\documentclass[prl, amsmath, onecolumn, amssymb, aps, superscriptaddress, floatfix, preprintnumbers, notitlepage, 10pt,reprint]{revtex4-1}
\usepackage{times}
\usepackage{graphicx}
\usepackage{amsmath, braket, amsfonts}
\usepackage{amssymb}
\usepackage{natbib}
\usepackage{todonotes}
\usepackage[modulo]{lineno}
\usepackage{bm, color, ulem}

\newcommand{\cyc}{\mathrm{cyc}}
\newcommand{\eff}{\mathrm{eff}}

\begin{document}

\title{Quantitative transport measurements of fractional quantum Hall energy gaps in edgeless graphene devices}
\author{H. Polshyn }
\affiliation{Department of Physics, University of California, Santa Barbara CA 93106 USA}
\author{H. Zhou}
\affiliation{Department of Physics, University of California, Santa Barbara CA 93106 USA}
\author{E. M. Spanton}
\affiliation{Department of Physics, University of California, Santa Barbara CA 93106 USA}
\author{T. Taniguchi}
\affiliation{Advanced Materials Laboratory, National Institute for Materials Science, Tsukuba, Ibaraki 305-0044, Japan}
\author{K. Watanabe}
\affiliation{Advanced Materials Laboratory, National Institute for Materials Science, Tsukuba, Ibaraki 305-0044, Japan}
\author{A. F. Young}
\affiliation{Department of Physics, University of California, Santa Barbara CA 93106 USA}
\begin{abstract}
Owing to their wide tunability, spin- and valley internal degrees of freedom, and low disorder, graphene heterostructures are emerging as a promising experimental platform for fractional quantum Hall (FQH) studies. Surprisingly, however, transport measurements reveal many fewer FQH states than bulk capacitive probes. Here, we report the fabrication of dual graphite-gated monolayer graphene devices in an edgeless Corbino-type geometry that showing deep FQH sequences. Thermal activation gaps reveal a tunable crossover between single- and multi-component FQH states in the zero energy Landau level, while the first Landau level is found to host an unexpected valley-ordered state at $\nu=-4$.
\end{abstract}

\maketitle


Advances in graphene sample fabrication over the past decade including suspension~\cite{bolotin_ultrahigh_2008,du_approaching_2008}, hexagonal boron nitride (hBN) gate dielectrics~\cite{dean_boron_2010}, hBN encapsulation~\cite{mayorov_micrometer-scale_2011,wang_one-dimensional_2013}, and most recently dual single-crystal graphite gates~\cite{zibrov_tunable_2017}, have increased sample quality in graphene heterostructures so that a variety of fractional quantum Hall~(FQH) states can now be accessed\cite{bolotin_observation_2009, du_fractional_2009, dean_multicomponent_2011, feldman_unconventional_2012, feldman_fractional_2013, kou_electron-hole_2014, ki_observation_2014, maher_tunable_2014, amet_composite_2015, bestwick_robust_2016, zibrov_tunable_2017, li_even_2017, zibrov_even_2017}. 
In contrast to III-V quantum wells, the exceptional quality of the sample bulk in the most recent generation of dual-graphite gated devices does not manifest clearly in transport measurements\cite{zibrov_tunable_2017,li_even_2017}, a discrepancy attributed to the uncontrolled electrostatics and chemistry of graphene crystal boundaries. Unfortunately, this has precluded quantitative studies of FQH energy gaps in ultra-clean devices, and poses an obstacle to realizing transport transport devices based on FQH edge states. 

In this Letter, we report the fabrication of dual-graphite gated, hBN encapsulated devices with Corbino topology. Because current flows only through the sample bulk, Corbino device enable direct measurement of the longitudinal conductivity $G_{xx}$, and are ideal for thermal activation gap measurements. However, previous realizations of Corbino geometry graphene heterostructures have lacked the requisite sample quality~\cite{Yan_2010, Zhu_2018, Peters_2014, Zhao_2012} to access the FQH regime.  Our fabrication technique, moreover, can be extended to a variety of other geometries including those incorporating electrostatically defined edges.

\begin{figure}[t!]
	\includegraphics[]{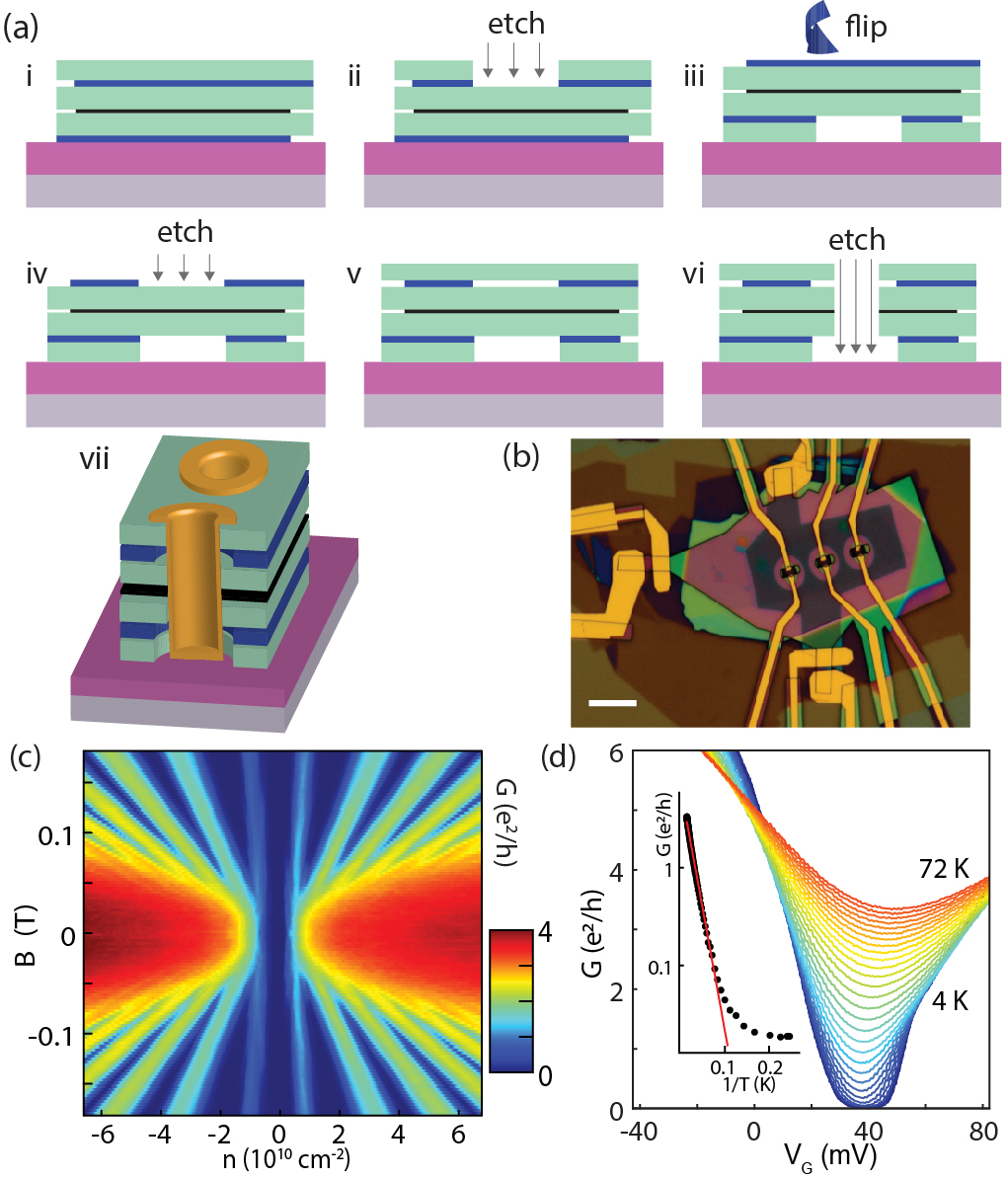}
\caption{Edgeless graphene devices.
(a) Steps for fabricating internal contacts.
(i) Dry transfer produces an hBN/graphite/hBN/graphene/hBN/graphite heterostructure.  (ii) A hole is etched in the top hBN and first graphite layer. (iii) The stack is flipped upside down, exposing the 2nd graphite layer.  (iv) The exposed graphite is etched and (v) another hBN flake deposited.  (vi) A hole is etched through the entire stack to expose the graphene for (viii) edge contacting\cite{wang_one-dimensional_2013}.
(b) Optical micrograph of Device A. Scale bar is 10~$\mu \text{m}$.
(c) Conductance of Device B at low magnetic fields.  The insulating state persisting through $B=0$ at charge neutrality is associated with broken AB sublattice symmetry~\cite{hunt_massive_2013,amet_insulating_2013}.
(d) Thermally activated transport at charge neutrality in device A. The measured activation gap $\Delta_{AB}\approx$114~K.}
\label{CorbinoDevice}
\vspace{-10pt}
\end{figure}

\begin{figure*}[!t]
\includegraphics[]{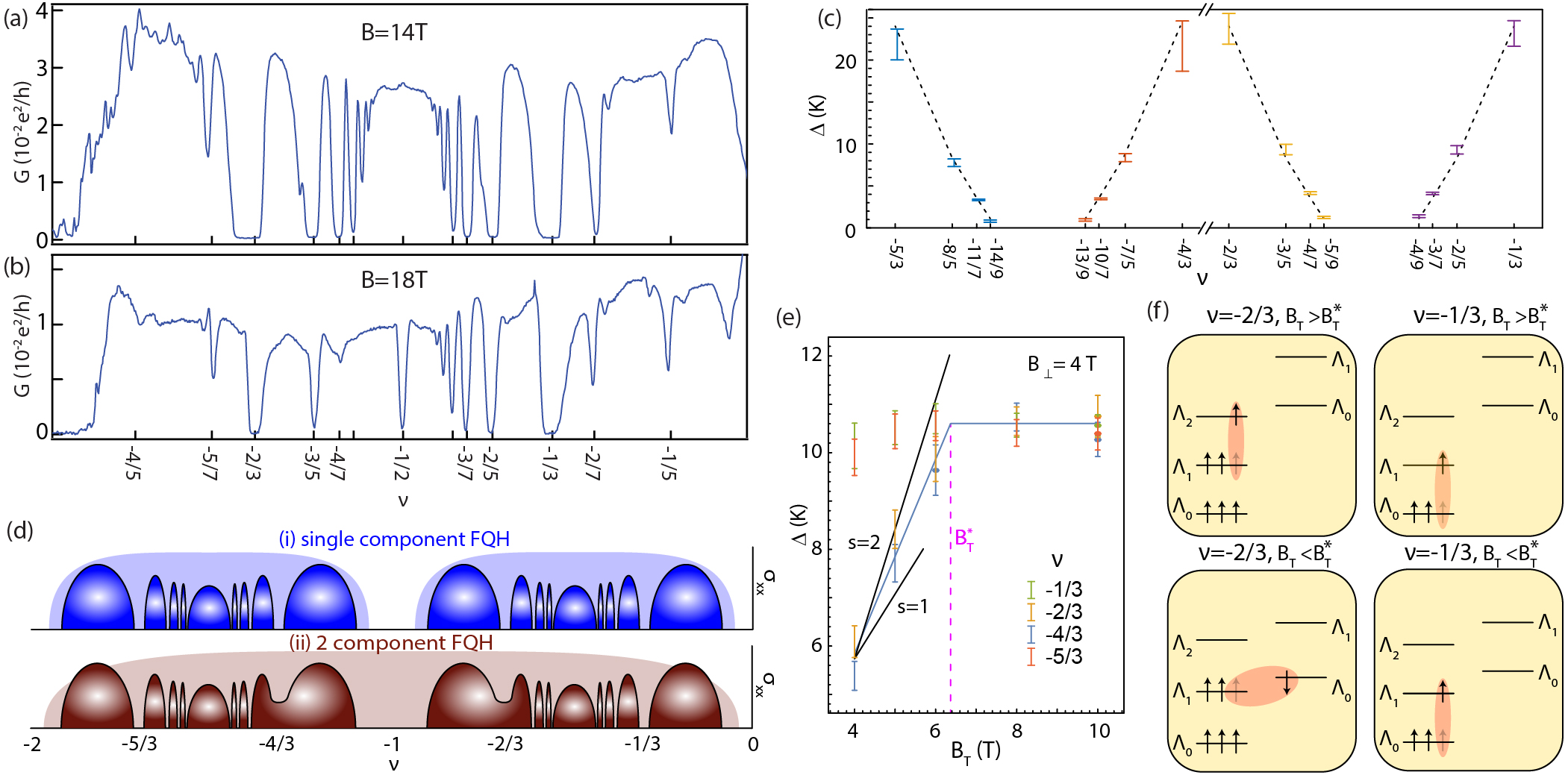}
\caption{
 FQH gaps in the zero energy LL. 
  (a) FQH states in Device A at $B_\perp=B_T=14$~T. Conductance minima are highly symmetric under particle hole inversion across the single LL, $\nu\leftrightarrow-1-\nu$.
 (b) FQH states in Device~A at 18T.
(c) FQH activation gaps in Device~A at $B_\perp/B_T$=10T/14T. Dashed lines are numerical results for a single component system~\cite{morf_excitation_2002} with  $\epsilon=\epsilon^{hBN}_\parallel=6.6$ and phenomenological broadening $\Gamma$=7.2K.   
(d) Schematic of particle hole symmetry in the case of single component (i) and 2-component FQH (ii). In the 2-component case, particle hole symmetry is present only across the whole doubly degenerate LL, while in the single component regime each singly degenerate LL is particle hole symmetric.
 (e) $B_T$ dependence of $n/3$ activation gaps for $B_{\perp}=4$~T.  The gaps at -2/3 and -4/3 grow with $B_{T}$ below $B_{T}^*\approx 6.4\text{~T}$ before saturating.  Black lines indicates expected slope for spin flip excitations involving $s=1$ and $s=2$ flipped spins.
 (f) $\Lambda$-level energy diagram for $\nu=-1/3$ and $\nu=-2/3$.  Left- and right-hand $\Lambda$-levels are  spin up and down states, respectively. For $B_T<B_T^*$, the polarized $-2/3$ state has low energy excitations consisting of a spin reverse particle-hole pair, while the $-1/3$ state admits only spinless particle-hole excitations. For $B_T>B_T^*$, particle-hole excitations prevail for all states.
\label{fig:N0}}
\end{figure*}

Our fabrication process (Fig. \ref{CorbinoDevice}(a)) begins with a dry-transfered van der Waals stack~\cite{wang_one-dimensional_2013} comprising a graphene sheet sandwiched between two graphite gates and two hBN spacer layer. We use lithography and a CHF$_3$/O$_2$ etch to shape the top graphite layer, which will become the bottom gate of the completed device.  The stack is then picked up using a PPC film, and the film removed from a carrier PDMS substrate and placed stack-side-up on a new chip. The underlying PPC film is then thermally sublimated at 375$^\circ$~C, leaving the inverted stack on the chip surface.  The stack is shaped with another CHF$_3$/O$_2$ etch, and covered with a fourth hBN flake to isolate the exposed edges of the graphite gates. A final etch shapes the device and opens internal apertures for Cr/Pd/Au contacts to the graphene flake. Flipping the stack allows independent shaping of the top and bottom gates, permitting a wide variety of geometries with internal contacts.

We focus on two samples with large substrate-induced sublattice splitting $\Delta_{AB}\approx $~10~meV~\cite{hunt_massive_2013,amet_insulating_2013}. This is evidenced by $B_\perp$ independent  insulating state at charge neutrality  that shows thermally activated behavior (Fig~\ref{CorbinoDevice}(c-d)).  Integer quantum Hall features emerge below 50 mT, while FQH features emerge starting at $B_\perp\approx$1T.
Fig.~\ref{fig:N0}(a) shows transport measurements at $B_\perp$=14~T at T$\approx$30~mK.  We observe a series of insulating states at fillings $\nu=\frac{p}{2p\pm1}$ with $p$ up to 7, as well as several states $\nu=\frac{p}{4p\pm1}$ sequence.  An insulating state at $\nu=-1/2$ appears for a narrow range of magnetic field around 18~ T (Fig.~\ref{fig:N0}(b)), consistent with recent capacitance measurements in devices with finite $\Delta_{AB}$~\cite{zibrov_even_2017}.

The high quality of the insulating FQH states and experimental access to the conductivity allow us to measure FQH gaps using thermal activation. The FQH activation gap $^\nu\Delta$ measures the energy of the lowest energy charged excitation at filling $\nu$, which for a single component system is a quasiparticle/quasihole pair.  In graphene quantum Hall systems, however, the spin and valley degrees of freedom allow for activated charge transport via low energy spin- or valley-textures, known as skyrmions.  Skyrmions manifested in early studies of graphene FQH effects as strong violations of particle hole symmetry across a single component LL. Measurements found robust charge gaps at $\nu=\pm$4/3, $\pm$2/3, and $\pm$1/3 but strong suppression of FQH at $\nu=\pm5/3$, attributed to  low energy valley skyrmions at the latter filling~\cite{dean_multicomponent_2011,feldman_unconventional_2012}.  
Spin and valley physics is further complicated by strong valley anisotropies near $\nu=0$~\cite{abanin_fractional_2013}, precluding quantitative comparison of FQH gaps to numerical calculations which are most accurate for systems without internal degeneracy.

\begin{figure}[!ht]
\includegraphics[]{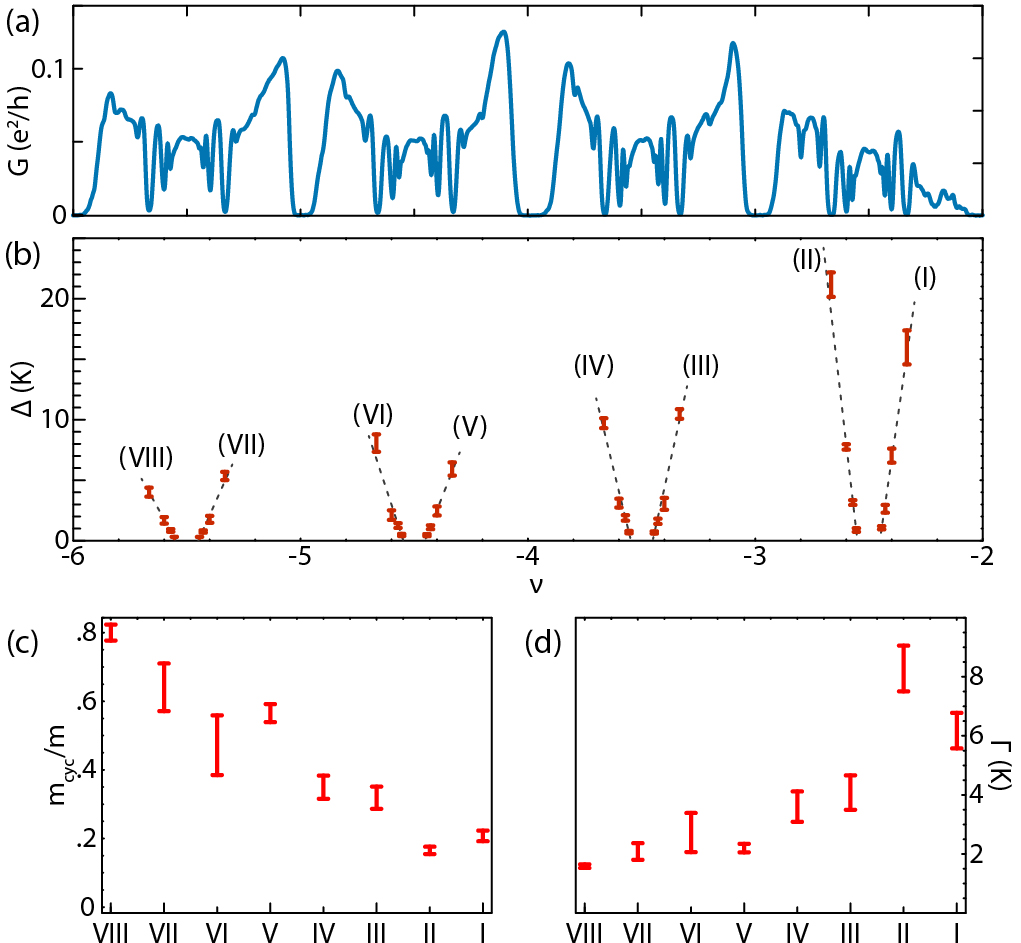}
\caption{Fractional and integer QH gaps in the first excited LL.
(a) Conductance of Devices~A measured at B = 10~T.
(b) Measured FQH activation gaps. Dashed lines are linear fits to the function $\Delta=e B_{\eff}/m_{\cyc}-\Gamma$ (defined in the main text) for each FQH series labeled by the numerals (I-VIII).
(c) Composite fermion cyclotron mass $m_{\cyc}$ and (d) broadening $\Gamma$ extracted from the linear fits for the different FQH series I-VIII in the first excited LL.}
\label{fig:N1}
\end{figure}

Fig.~\ref{fig:N0}(c) shows thermal activation gaps for a range of fractional fillings at perpendicular magnetic field $B_\perp=10$~T and total magnetic field $B_T=14$~T. In contrast to prior experiments\cite{dean_multicomponent_2011,feldman_unconventional_2012,zibrov_even_2017}, 
the energy gaps are highly particle hole symmetric across individual spin- and valley resolved LLs, as expected for single component FQH systems. 
Indeed, our measured FQH gaps are well matched to exact diagonalization calculations~\cite{morf_excitation_2002} using only a single phenomenological LL broadening parameter, $\Gamma$, to capture the effects of disorder, so that ${^\nu \Delta_{\textrm{meas}}}={^\nu\Delta_{ED}}-\Gamma$.  All four series of gaps within the lowest LL are well fit by $\Gamma=7.2$~K. For comparison, similar analysis on a GaAs 2DEG of mobility $7\times 10^6$~cm$^2$/V sec found $\Gamma=2$~K from fitting the behavior of the 1/3 state\cite{dethlefsen_signatures_2006_b}.
Restriction of FQH excitations to a single spin component is further supported by the absence of any dependence on $B_T$, which influences the energy of spin textured excitations via the  Zeeman energy $g \mu_B B_T$ (here g=2 is the electron gyromagnetic ratio, $\mu_B$ is the Bohr magneton). We find that gaps at $B_\perp=10$~T, $B_T=14$~T are equal to those measured at $B_\perp=B_T=10$~T within experimental error, consistent with the second spin branch remaining inert.  

Our results point to the critical role of $\Delta_{AB}$ in determining the excitation spectrum of FQH states in graphene.
The role of internal degeneracy in FQH systems is controlled by the ratio between the single-particle splitting of that degeneracy (the spin- or valley-Zeeman energy $E_Z$) and the Coulomb energy, $\kappa=E_Z/E_C$, where  $E_C=e^2/(\epsilon \ell_B)\approx 8.5 \mathrm{meV}\sqrt{B_\perp [\mathrm{T}]}$ using the measured~\cite{geick_normal_1966} in-plane dielectric constant of the encapsulating hBN layers $\epsilon=6.6$.
Theoretical calculations~\cite{balram_phase_2015} suggest that multicomponent physics is relevant in the FQH regime only for small values of $\kappa<0.05$, with the precise threshold strongly dependent on the filling factor.  In graphene, the spin and valley degeneracy can each play a role; however, in the lowest LL, the large AB-sublattice splitting in our devices amounts to a large valley Zeeman effect~\cite{hunt_massive_2013} so that $\kappa>0.275$ over the range $0<B_\perp<18$~T.  The valley degree of freedom is thus inert within the lowest LL.  Data in Fig.~\ref{fig:N0}(c) correspond to $\kappa=0.06$, consistent with single component behavior.  

However, the regime of low $\kappa$ is achievable in the spin sector, with $\kappa=0.027$ at $B_\perp=B_T=4$~T.  Indeed, at low magnetic field several gaps within the lowest LL show strong $B_T$ dependence. Figure~\ref{fig:N0}(e) shows the evolution with $B_T$ of the $\nu=-1/3,-2/3,-4/3$, and $-5/3$ gaps at $B_\perp=4$~T. While the $\nu=-1/3$ and $-5/3$ gaps are independent of in-plane magnetic field, the $\nu=-2/3$ and $-4/3$ gaps grow rapidly with $B_\perp$, saturating for $B_\perp\gtrsim B_T^*\approx 6.4$~T (equivalent to $\kappa\gtrsim 0.043$). Above $B_T^*$, all four gaps are equal within experimental error. $B_T$ thus tunes a transition between 2-component and single component FQH physics.

The behavior at all $n/3$ fillings is qualitatively captured by considering a simplified noninteracting composite fermion (CF) model, sketched in Fig.~\ref{fig:N0}(f). In the CF picture, FQH states at filling $\nu=p/(2p\pm1)$ and external field $B$ are envisioned as integer quantum Hall effects of emergent, \textit{noninteracting} composite particles consisting of an electron and two magnetic flux quanta at an effective magnetic field of $B_{\eff}=1-2\lfloor\nu\rfloor$ and filling factor $\nu_{\eff}=p$. Thus the $-1/3$ state corresponds to filling a single CF LL, known as a $\Lambda$-level, while the $-2/3$ consists of filling 2 $\Lambda$ levels.  The $-5/3$ ($-4/3$) state is related to the $-1/3$ ($-2/3$) state by particle hole symmetry across the 2-component LL, $\nu\leftrightarrow-2-\nu$.

In this picture, the Zeeman energy at $B_\perp/B_T=4T/4T$ is sufficient to spin polarize the $-2/3$ ground state, but small enough that the low energy excitations are spin flips~\cite{mandal_low-energy_2001,wurstbauer_observation_2011}. For $B_T>B_T^*$, however, the increased Zeeman energy makes the spin flip excitation energetically unfavorable, resulting in a crossover to a conventional inter-$\Lambda$ level particle-hole excitation without a reversed spin.  At $\nu=-1/3$, in contrast, spin-flip excitations are not favored even at the lowest values of $\kappa$ probed.

In real systems, residual composite fermion interactions lead to significant corrections to energy gaps and the details of spin transitions, which can be captured by numerical simulations.  The absence of spin-flip excitations at $\nu=-1/3$ is consistent with such calculations, which predict such excitations for $\kappa<0.009$, or $B_\perp=B_T<0.44$~T in hBN encapsulated graphene~\cite{balram_fractionally_2015}.  In an interacting composite fermion picture, moreover, the spin flip excitations themselves can involve multiple spins, which manifest in the $B_T$ dependence of energy gaps as $\partial \Delta/\partial B_T=s g\mu_B$ where $s$ corresponds to the number of flipped spins.  Our observation of $s>1$ (see Fig.~\ref{fig:N0}(e)) suggests that excitations at $\nu=-2/3, -4/3$ are extended spin textures rather than single reversed spins. The nature of thermal spin excitations at $\nu=2/3$ has only begun to be addressed numerically~\cite{vyborny_charge-spin_2009}, however.  Finally, in the absence of a Zeeman effect, interactions favor a spin unpolarized 2/3 state for $\kappa<0.017$~\cite{balram_phase_2015,geraedts_competing_2015}.  A spin unpolarized $-2/3$ is thus expected at $B_\perp=B_T<1.8$~T, just below the regime where the 2/3 state is develops in our samples. 

The first excited LL (-6$<\nu<$-2) also shows robust FQH sequences (Fig.~\ref{fig:N1}(a)). Activation gaps, although similarly Zeeman energy independent (see Fig.~S3(c)-(d)~\cite{SI}), diverge sharply from those in lowest LL (Fig.~\ref{fig:N0}(c)).  Most prominently, the level is strongly particle-hole asymmetryic; i.e., the energy gap $^{-2-\nu}\Delta\neq ^{-6+\nu}\Delta$.  This asymmetry indicates that LL mixing plays an important role in determining the size of activation gaps. In this picture, FQH states in the first excited LL mix at high $|\nu|$ mix more heavily with the higher LL, whose orbital structure is less favorable to FQH states.  Because applicable numerical simulations are not available, we analyze the data using a noninteracting composite fermion picture. The CF picture predicts a linear dependence of the energy gaps on $\nu$ within each FQH series, a trend well matched by the data, and allows us to quantify trends in $\nu$ across the level.  In addition to the broadening $\Gamma$ defined above, linear fits are parameterized by a phenomenological composite fermion cyclotron mass $m_{\cyc}$ such that $\Delta_{meas}=\frac{\hbar e B_{\eff}}{m_{\cyc}}-\Gamma$.  Figs.~\ref{fig:N1}(c)-(d) show the result of free fits of the measured gaps across the LL.

\begin{figure}[t!]
\includegraphics[]{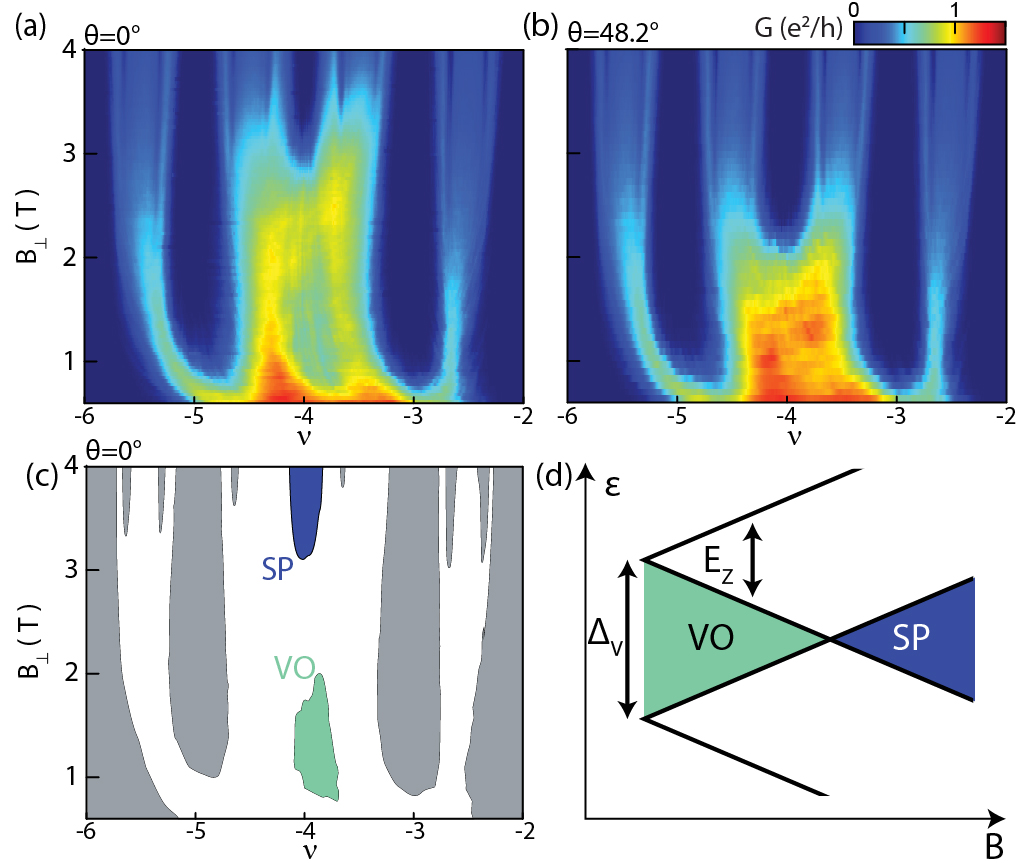}
\caption{Valley ordered phase at $\nu=-4$ in Device~B (for Device A see Fig.~S3(e)~\cite{SI}).
(a) Conductance in the first excited Landau level at $B_\perp=B_T$.  Two insulating states are visible, at low- and high $B_\perp$, noted schematically in (b).
(c) Conductance in the first excited LL in tilted field, with $B_\perp=0.67 B_T$.  The low $B_\perp$ state is suppressed.
(d) Level crossing model for the phase transition.  A valley ordered state (VO) driven by a valley splitting $\Delta_V$ (which may have either single-particle or many body origin) is suppressed by the Zeeman splitting $E_Z$, which favors a spin-polarized state (SP).  }
\label{fig:valleyOrder}
\end{figure}

We also find a new phase at integer filling $\nu=-4$, corresponding to half filling of the first excited LL.  Fig.~\ref{fig:valleyOrder}(a)-(b) shows low $B_\perp$ measurements in the first excited LL, with a phase transition at $B_\perp\approx2.3T$ evident as a rise in conductivity at $\nu=-4$ separating distinct low- and high-B insulating states. Increasing $B_T$ by tilting the field strengthens the high $B_\perp$ state, suggesting that it is spin polarized (SP) while the low $B_\perp$ insulator is a spin unpolarized, and consequently valley ordered (VO), quantum Hall ferromagnetic state (Fig.~\ref{fig:valleyOrder}(c)).  The transition can be understood phenomenologically by competition between the spin Zeeman effect and a valley-splitting $\Delta_V$, with the transition occurring when $\Delta_V=E_Z$, allowing us to estimate $\Delta_V\approx 3K$ (Fig.~\ref{fig:valleyOrder}(d)).  

The origin of $\Delta_V$ is not clear.  While a sublattice gap $\Delta_{AB}$ generates a large single particle splitting in the lowest LL, it generates a much smaller splitting in the first excited LL of $\Delta_V\ll 1K$.  It is thus instructive to compare the competition between the phases in the  first  half-filled LL ($\nu=-4$) in which a valley Zeeman due to the band structure is weak, with the competition between phases in the  lowest half-filled LL ($\nu=0$) in samples with $\Delta_{AB}\approx 0$.
In the latter case, anisotropy of the Coulomb interactions themselves~\cite{alicea_graphene_2006_b,kharitonov_phase_2012} choose between a set of nearly degenerate states polarized  within the spin- and valley- isospin space. The resulting ground state is believed to be an antiferromagnet, which can be suppressed in favor of a spin polarized state by large $B_T$.  Determining the nature of the low-$B$ insulating state at $\nu=-4$ may involve a similarly subtle interplay of anisotropies, with $\Delta_V$ having either single-particle or many-body origin.

In conclusion, we have introduced a versatile fabrication method for producing van der Waals heterostructure devices in which measured transport occurs entirely through the sample bulk.  Precision measurements of activated transport reveal that graphene 2D electron systems are comparable to high mobility GaAs quantum wells in sample quality, and can realize much of the same fractional quantum Hall physics.  Our results clarify the role of spin physics in graphene FQH states, which have been proposed a a basis for topological superconductivity using superconducting proximity effects in the fractional quantum Hall regime~\cite{alicea_topological_2016}.  They further reveal a new, likely correlation driven insulating phase at $\nu=-4$, which will be the basis for future study of correlated electrons in a magnetic field.  Future experiments may also leverage this fabrication technique, for example to study edge transport in entirely gate-defined devices fabricated in the interior of a single graphene flake.

\begin{acknowledgments}
We thank Cory Dean, S. Chen, Y. Zeng, M. Yankowitz and J. Li for discussing their unpublished data and  for sharing the stack inversion technique. 
The authors acknowledge further discussions of the results with I. Sodemann, M. Zaletel, C. Nayak, and J. Jain.
AFY, HP, HZ, and EMS were supported by the ARO under awards 69188PHH and MURI W911NF-17-1-0323.
A portion of this work was performed at the National High Magnetic Field Laboratory, which is supported by National Science Foundation Cooperative Agreement No. DMR-1644779 and the State of Florida.
K.W. and T.T. acknowledge support from the Elemental Strategy Initiative conducted by the MEXT, Japan and JSPS KAKENHI grant number JP15K21722. AFY acknowledges the support of the David and Lucile Packard Foundation.
EMS acknowledges the support of the Elings Prize Fellowship in Science of the California Nanosystems Institute at the University of California, Santa Barbara.  
\end{acknowledgments}


\begin{thebibliography}{38}%
\makeatletter
\providecommand \@ifxundefined [1]{%
 \@ifx{#1\undefined}
}%
\providecommand \@ifnum [1]{%
 \ifnum #1\expandafter \@firstoftwo
 \else \expandafter \@secondoftwo
 \fi
}%
\providecommand \@ifx [1]{%
 \ifx #1\expandafter \@firstoftwo
 \else \expandafter \@secondoftwo
 \fi
}%
\providecommand \natexlab [1]{#1}%
\providecommand \enquote  [1]{``#1''}%
\providecommand \bibnamefont  [1]{#1}%
\providecommand \bibfnamefont [1]{#1}%
\providecommand \citenamefont [1]{#1}%
\providecommand \href@noop [0]{\@secondoftwo}%
\providecommand \href [0]{\begingroup \@sanitize@url \@href}%
\providecommand \@href[1]{\@@startlink{#1}\@@href}%
\providecommand \@@href[1]{\endgroup#1\@@endlink}%
\providecommand \@sanitize@url [0]{\catcode `\\12\catcode `\$12\catcode
  `\&12\catcode `\#12\catcode `\^12\catcode `\_12\catcode `\%12\relax}%
\providecommand \@@startlink[1]{}%
\providecommand \@@endlink[0]{}%
\providecommand \url  [0]{\begingroup\@sanitize@url \@url }%
\providecommand \@url [1]{\endgroup\@href {#1}{\urlprefix }}%
\providecommand \urlprefix  [0]{URL }%
\providecommand \Eprint [0]{\href }%
\providecommand \doibase [0]{http://dx.doi.org/}%
\providecommand \selectlanguage [0]{\@gobble}%
\providecommand \bibinfo  [0]{\@secondoftwo}%
\providecommand \bibfield  [0]{\@secondoftwo}%
\providecommand \translation [1]{[#1]}%
\providecommand \BibitemOpen [0]{}%
\providecommand \bibitemStop [0]{}%
\providecommand \bibitemNoStop [0]{.\EOS\space}%
\providecommand \EOS [0]{\spacefactor3000\relax}%
\providecommand \BibitemShut  [1]{\csname bibitem#1\endcsname}%
\let\auto@bib@innerbib\@empty
\bibitem [{\citenamefont {Bolotin}\ \emph {et~al.}(2008)\citenamefont
  {Bolotin}, \citenamefont {Sikes}, \citenamefont {Jiang}, \citenamefont
  {Klima}, \citenamefont {Fudenberg}, \citenamefont {Hone}, \citenamefont
  {Kim},\ and\ \citenamefont {Stormer}}]{bolotin_ultrahigh_2008}%
  \BibitemOpen
  \bibfield  {author} {\bibinfo {author} {\bibfnamefont {K.~I.}\ \bibnamefont
  {Bolotin}}, \bibinfo {author} {\bibfnamefont {K.~J.}\ \bibnamefont {Sikes}},
  \bibinfo {author} {\bibfnamefont {Z.}~\bibnamefont {Jiang}}, \bibinfo
  {author} {\bibfnamefont {M.}~\bibnamefont {Klima}}, \bibinfo {author}
  {\bibfnamefont {G.}~\bibnamefont {Fudenberg}}, \bibinfo {author}
  {\bibfnamefont {J.}~\bibnamefont {Hone}}, \bibinfo {author} {\bibfnamefont
  {P.}~\bibnamefont {Kim}}, \ and\ \bibinfo {author} {\bibfnamefont {H.~L.}\
  \bibnamefont {Stormer}},\ }\href@noop {} {\bibfield  {journal} {\bibinfo
  {journal} {Solid State Communications}\ }\textbf {\bibinfo {volume} {146,}},\
  \bibinfo {pages} {351} (\bibinfo {year} {2008})}\BibitemShut {NoStop}%
\bibitem [{\citenamefont {Du}\ \emph {et~al.}(2008)\citenamefont {Du},
  \citenamefont {Skachko}, \citenamefont {Barker},\ and\ \citenamefont
  {Andrei}}]{du_approaching_2008}%
  \BibitemOpen
  \bibfield  {author} {\bibinfo {author} {\bibfnamefont {X.}~\bibnamefont
  {Du}}, \bibinfo {author} {\bibfnamefont {I.}~\bibnamefont {Skachko}},
  \bibinfo {author} {\bibfnamefont {A.}~\bibnamefont {Barker}}, \ and\ \bibinfo
  {author} {\bibfnamefont {E.~Y.}\ \bibnamefont {Andrei}},\ }\href
  {http://dx.doi.org/10.1038/nnano.2008.199} {\bibfield  {journal} {\bibinfo
  {journal} {Nature Nanotechnology}\ }\textbf {\bibinfo {volume} {3}},\
  \bibinfo {pages} {491} (\bibinfo {year} {2008})}\BibitemShut {NoStop}%
\bibitem [{\citenamefont {Dean}\ \emph {et~al.}(2010)\citenamefont {Dean},
  \citenamefont {Young}, \citenamefont {Meric}, \citenamefont {Lee},
  \citenamefont {Wang}, \citenamefont {Sorgenfrei}, \citenamefont {Watanabe},
  \citenamefont {Taniguchi}, \citenamefont {Kim}, \citenamefont {Shepard},\
  and\ \citenamefont {Hone}}]{dean_boron_2010}%
  \BibitemOpen
  \bibfield  {author} {\bibinfo {author} {\bibfnamefont {C.~R.}\ \bibnamefont
  {Dean}}, \bibinfo {author} {\bibfnamefont {A.~F.}\ \bibnamefont {Young}},
  \bibinfo {author} {\bibfnamefont {I.}~\bibnamefont {Meric}}, \bibinfo
  {author} {\bibfnamefont {C.}~\bibnamefont {Lee}}, \bibinfo {author}
  {\bibfnamefont {L.}~\bibnamefont {Wang}}, \bibinfo {author} {\bibfnamefont
  {S.}~\bibnamefont {Sorgenfrei}}, \bibinfo {author} {\bibfnamefont
  {K.}~\bibnamefont {Watanabe}}, \bibinfo {author} {\bibfnamefont
  {T.}~\bibnamefont {Taniguchi}}, \bibinfo {author} {\bibfnamefont
  {P.}~\bibnamefont {Kim}}, \bibinfo {author} {\bibfnamefont {K.~L.}\
  \bibnamefont {Shepard}}, \ and\ \bibinfo {author} {\bibfnamefont
  {J.}~\bibnamefont {Hone}},\ }\href {http://dx.doi.org/10.1038/nnano.2010.172}
  {\bibfield  {journal} {\bibinfo  {journal} {Nature Nanotechnology}\ }\textbf
  {\bibinfo {volume} {5}},\ \bibinfo {pages} {722} (\bibinfo {year}
  {2010})}\BibitemShut {NoStop}%
\bibitem [{\citenamefont {Mayorov}\ \emph {et~al.}(2011)\citenamefont
  {Mayorov}, \citenamefont {Gorbachev}, \citenamefont {Morozov}, \citenamefont
  {Britnell}, \citenamefont {Jalil}, \citenamefont {Ponomarenko}, \citenamefont
  {Blake}, \citenamefont {Novoselov}, \citenamefont {Watanabe}, \citenamefont
  {Taniguchi},\ and\ \citenamefont {Geim}}]{mayorov_micrometer-scale_2011}%
  \BibitemOpen
  \bibfield  {author} {\bibinfo {author} {\bibfnamefont {A.~S.}\ \bibnamefont
  {Mayorov}}, \bibinfo {author} {\bibfnamefont {R.~V.}\ \bibnamefont
  {Gorbachev}}, \bibinfo {author} {\bibfnamefont {S.~V.}\ \bibnamefont
  {Morozov}}, \bibinfo {author} {\bibfnamefont {L.}~\bibnamefont {Britnell}},
  \bibinfo {author} {\bibfnamefont {R.}~\bibnamefont {Jalil}}, \bibinfo
  {author} {\bibfnamefont {L.~A.}\ \bibnamefont {Ponomarenko}}, \bibinfo
  {author} {\bibfnamefont {P.}~\bibnamefont {Blake}}, \bibinfo {author}
  {\bibfnamefont {K.~S.}\ \bibnamefont {Novoselov}}, \bibinfo {author}
  {\bibfnamefont {K.}~\bibnamefont {Watanabe}}, \bibinfo {author}
  {\bibfnamefont {T.}~\bibnamefont {Taniguchi}}, \ and\ \bibinfo {author}
  {\bibfnamefont {A.~K.}\ \bibnamefont {Geim}},\ }\href
  {http://dx.doi.org/10.1021/nl200758b} {\bibfield  {journal} {\bibinfo
  {journal} {Nano Letters}\ }\textbf {\bibinfo {volume} {11}},\ \bibinfo
  {pages} {2396} (\bibinfo {year} {2011})}\BibitemShut {NoStop}%
\bibitem [{\citenamefont {Wang}\ \emph {et~al.}(2013)\citenamefont {Wang},
  \citenamefont {Meric}, \citenamefont {Huang}, \citenamefont {Gao},
  \citenamefont {Gao}, \citenamefont {Tran}, \citenamefont {Taniguchi},
  \citenamefont {Watanabe}, \citenamefont {Campos}, \citenamefont {Muller},
  \citenamefont {Guo}, \citenamefont {Kim}, \citenamefont {Hone}, \citenamefont
  {Shepard},\ and\ \citenamefont {Dean}}]{wang_one-dimensional_2013}%
  \BibitemOpen
  \bibfield  {author} {\bibinfo {author} {\bibfnamefont {L.}~\bibnamefont
  {Wang}}, \bibinfo {author} {\bibfnamefont {I.}~\bibnamefont {Meric}},
  \bibinfo {author} {\bibfnamefont {P.~Y.}\ \bibnamefont {Huang}}, \bibinfo
  {author} {\bibfnamefont {Q.}~\bibnamefont {Gao}}, \bibinfo {author}
  {\bibfnamefont {Y.}~\bibnamefont {Gao}}, \bibinfo {author} {\bibfnamefont
  {H.}~\bibnamefont {Tran}}, \bibinfo {author} {\bibfnamefont {T.}~\bibnamefont
  {Taniguchi}}, \bibinfo {author} {\bibfnamefont {K.}~\bibnamefont {Watanabe}},
  \bibinfo {author} {\bibfnamefont {L.~M.}\ \bibnamefont {Campos}}, \bibinfo
  {author} {\bibfnamefont {D.~A.}\ \bibnamefont {Muller}}, \bibinfo {author}
  {\bibfnamefont {J.}~\bibnamefont {Guo}}, \bibinfo {author} {\bibfnamefont
  {P.}~\bibnamefont {Kim}}, \bibinfo {author} {\bibfnamefont {J.}~\bibnamefont
  {Hone}}, \bibinfo {author} {\bibfnamefont {K.~L.}\ \bibnamefont {Shepard}}, \
  and\ \bibinfo {author} {\bibfnamefont {C.~R.}\ \bibnamefont {Dean}},\ }\href
  {http://www.sciencemag.org/content/342/6158/614} {\bibfield  {journal}
  {\bibinfo  {journal} {Science}\ }\textbf {\bibinfo {volume} {342}},\ \bibinfo
  {pages} {614} (\bibinfo {year} {2013})}\BibitemShut {NoStop}%
\bibitem [{\citenamefont {Zibrov}\ \emph
  {et~al.}(2017{\natexlab{a}})\citenamefont {Zibrov}, \citenamefont {Kometter},
  \citenamefont {Zhou}, \citenamefont {Spanton}, \citenamefont {Taniguchi},
  \citenamefont {Watanabe}, \citenamefont {Zaletel},\ and\ \citenamefont
  {Young}}]{zibrov_tunable_2017}%
  \BibitemOpen
  \bibfield  {author} {\bibinfo {author} {\bibfnamefont {A.~A.}\ \bibnamefont
  {Zibrov}}, \bibinfo {author} {\bibfnamefont {C.}~\bibnamefont {Kometter}},
  \bibinfo {author} {\bibfnamefont {H.}~\bibnamefont {Zhou}}, \bibinfo {author}
  {\bibfnamefont {E.~M.}\ \bibnamefont {Spanton}}, \bibinfo {author}
  {\bibfnamefont {T.}~\bibnamefont {Taniguchi}}, \bibinfo {author}
  {\bibfnamefont {K.}~\bibnamefont {Watanabe}}, \bibinfo {author}
  {\bibfnamefont {M.~P.}\ \bibnamefont {Zaletel}}, \ and\ \bibinfo {author}
  {\bibfnamefont {A.~F.}\ \bibnamefont {Young}},\ }\href {\doibase
  10.1038/nature23893} {\bibfield  {journal} {\bibinfo  {journal} {Nature}\
  }\textbf {\bibinfo {volume} {549}},\ \bibinfo {pages} {360} (\bibinfo {year}
  {2017}{\natexlab{a}})}\BibitemShut {NoStop}%
\bibitem [{\citenamefont {Bolotin}\ \emph {et~al.}(2009)\citenamefont
  {Bolotin}, \citenamefont {Ghahari}, \citenamefont {Shulman}, \citenamefont
  {Stormer},\ and\ \citenamefont {Kim}}]{bolotin_observation_2009}%
  \BibitemOpen
  \bibfield  {author} {\bibinfo {author} {\bibfnamefont {K.~I.}\ \bibnamefont
  {Bolotin}}, \bibinfo {author} {\bibfnamefont {F.}~\bibnamefont {Ghahari}},
  \bibinfo {author} {\bibfnamefont {M.~D.}\ \bibnamefont {Shulman}}, \bibinfo
  {author} {\bibfnamefont {H.~L.}\ \bibnamefont {Stormer}}, \ and\ \bibinfo
  {author} {\bibfnamefont {P.}~\bibnamefont {Kim}},\ }\href
  {http://dx.doi.org/10.1038/nature08582} {\bibfield  {journal} {\bibinfo
  {journal} {Nature}\ }\textbf {\bibinfo {volume} {462}},\ \bibinfo {pages}
  {196} (\bibinfo {year} {2009})}\BibitemShut {NoStop}%
\bibitem [{\citenamefont {Du}\ \emph {et~al.}(2009)\citenamefont {Du},
  \citenamefont {Skachko}, \citenamefont {Duerr}, \citenamefont {Luican},\ and\
  \citenamefont {Andrei}}]{du_fractional_2009}%
  \BibitemOpen
  \bibfield  {author} {\bibinfo {author} {\bibfnamefont {X.}~\bibnamefont
  {Du}}, \bibinfo {author} {\bibfnamefont {I.}~\bibnamefont {Skachko}},
  \bibinfo {author} {\bibfnamefont {F.}~\bibnamefont {Duerr}}, \bibinfo
  {author} {\bibfnamefont {A.}~\bibnamefont {Luican}}, \ and\ \bibinfo {author}
  {\bibfnamefont {E.~Y.}\ \bibnamefont {Andrei}},\ }\href
  {http://dx.doi.org/10.1038/nature08522} {\bibfield  {journal} {\bibinfo
  {journal} {Nature}\ }\textbf {\bibinfo {volume} {462}},\ \bibinfo {pages}
  {192} (\bibinfo {year} {2009})}\BibitemShut {NoStop}%
\bibitem [{\citenamefont {Dean}\ \emph {et~al.}(2011)\citenamefont {Dean},
  \citenamefont {Young}, \citenamefont {Cadden-Zimansky}, \citenamefont {Wang},
  \citenamefont {Ren}, \citenamefont {Watanabe}, \citenamefont {Taniguchi},
  \citenamefont {Kim}, \citenamefont {Hone},\ and\ \citenamefont
  {Shepard}}]{dean_multicomponent_2011}%
  \BibitemOpen
  \bibfield  {author} {\bibinfo {author} {\bibfnamefont {C.~R.}\ \bibnamefont
  {Dean}}, \bibinfo {author} {\bibfnamefont {A.~F.}\ \bibnamefont {Young}},
  \bibinfo {author} {\bibfnamefont {P.}~\bibnamefont {Cadden-Zimansky}},
  \bibinfo {author} {\bibfnamefont {L.}~\bibnamefont {Wang}}, \bibinfo {author}
  {\bibfnamefont {H.}~\bibnamefont {Ren}}, \bibinfo {author} {\bibfnamefont
  {K.}~\bibnamefont {Watanabe}}, \bibinfo {author} {\bibfnamefont
  {T.}~\bibnamefont {Taniguchi}}, \bibinfo {author} {\bibfnamefont
  {P.}~\bibnamefont {Kim}}, \bibinfo {author} {\bibfnamefont {J.}~\bibnamefont
  {Hone}}, \ and\ \bibinfo {author} {\bibfnamefont {K.~L.}\ \bibnamefont
  {Shepard}},\ }\href {http://dx.doi.org/10.1038/nphys2007} {\bibfield
  {journal} {\bibinfo  {journal} {Nature Physics}\ }\textbf {\bibinfo {volume}
  {7}},\ \bibinfo {pages} {693} (\bibinfo {year} {2011})}\BibitemShut {NoStop}%
\bibitem [{\citenamefont {Feldman}\ \emph {et~al.}(2012)\citenamefont
  {Feldman}, \citenamefont {Krauss}, \citenamefont {Smet},\ and\ \citenamefont
  {Yacoby}}]{feldman_unconventional_2012}%
  \BibitemOpen
  \bibfield  {author} {\bibinfo {author} {\bibfnamefont {B.~E.}\ \bibnamefont
  {Feldman}}, \bibinfo {author} {\bibfnamefont {B.}~\bibnamefont {Krauss}},
  \bibinfo {author} {\bibfnamefont {J.~H.}\ \bibnamefont {Smet}}, \ and\
  \bibinfo {author} {\bibfnamefont {A.}~\bibnamefont {Yacoby}},\ }\href
  {\doibase 10.1126/science.1224784} {\bibfield  {journal} {\bibinfo  {journal}
  {Science}\ }\textbf {\bibinfo {volume} {337}},\ \bibinfo {pages} {1196}
  (\bibinfo {year} {2012})}\BibitemShut {NoStop}%
\bibitem [{\citenamefont {Feldman}\ \emph {et~al.}(2013)\citenamefont
  {Feldman}, \citenamefont {Levin}, \citenamefont {Krauss}, \citenamefont
  {Abanin}, \citenamefont {Halperin}, \citenamefont {Smet},\ and\ \citenamefont
  {Yacoby}}]{feldman_fractional_2013}%
  \BibitemOpen
  \bibfield  {author} {\bibinfo {author} {\bibfnamefont {B.~E.}\ \bibnamefont
  {Feldman}}, \bibinfo {author} {\bibfnamefont {A.~J.}\ \bibnamefont {Levin}},
  \bibinfo {author} {\bibfnamefont {B.}~\bibnamefont {Krauss}}, \bibinfo
  {author} {\bibfnamefont {D.~A.}\ \bibnamefont {Abanin}}, \bibinfo {author}
  {\bibfnamefont {B.~I.}\ \bibnamefont {Halperin}}, \bibinfo {author}
  {\bibfnamefont {J.~H.}\ \bibnamefont {Smet}}, \ and\ \bibinfo {author}
  {\bibfnamefont {A.}~\bibnamefont {Yacoby}},\ }\href {\doibase
  10.1103/PhysRevLett.111.076802} {\bibfield  {journal} {\bibinfo  {journal}
  {Physical Review Letters}\ }\textbf {\bibinfo {volume} {111}},\ \bibinfo
  {pages} {076802} (\bibinfo {year} {2013})}\BibitemShut {NoStop}%
\bibitem [{\citenamefont {Kou}\ \emph {et~al.}(2014)\citenamefont {Kou},
  \citenamefont {Feldman}, \citenamefont {Levin}, \citenamefont {Halperin},
  \citenamefont {Watanabe}, \citenamefont {Taniguchi},\ and\ \citenamefont
  {Yacoby}}]{kou_electron-hole_2014}%
  \BibitemOpen
  \bibfield  {author} {\bibinfo {author} {\bibfnamefont {A.}~\bibnamefont
  {Kou}}, \bibinfo {author} {\bibfnamefont {B.~E.}\ \bibnamefont {Feldman}},
  \bibinfo {author} {\bibfnamefont {A.~J.}\ \bibnamefont {Levin}}, \bibinfo
  {author} {\bibfnamefont {B.~I.}\ \bibnamefont {Halperin}}, \bibinfo {author}
  {\bibfnamefont {K.}~\bibnamefont {Watanabe}}, \bibinfo {author}
  {\bibfnamefont {T.}~\bibnamefont {Taniguchi}}, \ and\ \bibinfo {author}
  {\bibfnamefont {A.}~\bibnamefont {Yacoby}},\ }\href {\doibase
  10.1126/science.1250270} {\bibfield  {journal} {\bibinfo  {journal}
  {Science}\ }\textbf {\bibinfo {volume} {345}},\ \bibinfo {pages} {55}
  (\bibinfo {year} {2014})}\BibitemShut {NoStop}%
\bibitem [{\citenamefont {Ki}\ \emph {et~al.}(2014)\citenamefont {Ki},
  \citenamefont {Fal{\textquoteright}ko}, \citenamefont {Abanin},\ and\
  \citenamefont {Morpurgo}}]{ki_observation_2014}%
  \BibitemOpen
  \bibfield  {author} {\bibinfo {author} {\bibfnamefont {D.-K.}\ \bibnamefont
  {Ki}}, \bibinfo {author} {\bibfnamefont {V.~I.}\ \bibnamefont
  {Fal{\textquoteright}ko}}, \bibinfo {author} {\bibfnamefont {D.~A.}\
  \bibnamefont {Abanin}}, \ and\ \bibinfo {author} {\bibfnamefont {A.~F.}\
  \bibnamefont {Morpurgo}},\ }\href {\doibase 10.1021/nl5003922} {\bibfield
  {journal} {\bibinfo  {journal} {Nano Letters}\ }\textbf {\bibinfo {volume}
  {14}},\ \bibinfo {pages} {2135} (\bibinfo {year} {2014})}\BibitemShut
  {NoStop}%
\bibitem [{\citenamefont {Maher}\ \emph {et~al.}(2014)\citenamefont {Maher},
  \citenamefont {Wang}, \citenamefont {Gao}, \citenamefont {Forsythe},
  \citenamefont {Taniguchi}, \citenamefont {Watanabe}, \citenamefont {Abanin},
  \citenamefont {Papic}, \citenamefont {Cadden-Zimansky}, \citenamefont {Hone},
  \citenamefont {Kim},\ and\ \citenamefont {Dean}}]{maher_tunable_2014}%
  \BibitemOpen
  \bibfield  {author} {\bibinfo {author} {\bibfnamefont {P.}~\bibnamefont
  {Maher}}, \bibinfo {author} {\bibfnamefont {L.}~\bibnamefont {Wang}},
  \bibinfo {author} {\bibfnamefont {Y.}~\bibnamefont {Gao}}, \bibinfo {author}
  {\bibfnamefont {C.}~\bibnamefont {Forsythe}}, \bibinfo {author}
  {\bibfnamefont {T.}~\bibnamefont {Taniguchi}}, \bibinfo {author}
  {\bibfnamefont {K.}~\bibnamefont {Watanabe}}, \bibinfo {author}
  {\bibfnamefont {D.}~\bibnamefont {Abanin}}, \bibinfo {author} {\bibfnamefont
  {Z.}~\bibnamefont {Papic}}, \bibinfo {author} {\bibfnamefont
  {P.}~\bibnamefont {Cadden-Zimansky}}, \bibinfo {author} {\bibfnamefont
  {J.}~\bibnamefont {Hone}}, \bibinfo {author} {\bibfnamefont {P.}~\bibnamefont
  {Kim}}, \ and\ \bibinfo {author} {\bibfnamefont {C.~R.}\ \bibnamefont
  {Dean}},\ }\href {\doibase 10.1126/science.1252875} {\bibfield  {journal}
  {\bibinfo  {journal} {Science}\ }\textbf {\bibinfo {volume} {345}},\ \bibinfo
  {pages} {61} (\bibinfo {year} {2014})}\BibitemShut {NoStop}%
\bibitem [{\citenamefont {Amet}\ \emph {et~al.}(2015)\citenamefont {Amet},
  \citenamefont {Bestwick}, \citenamefont {Williams}, \citenamefont {Balicas},
  \citenamefont {Watanabe}, \citenamefont {Taniguchi},\ and\ \citenamefont
  {Goldhaber-Gordon}}]{amet_composite_2015}%
  \BibitemOpen
  \bibfield  {author} {\bibinfo {author} {\bibfnamefont {F.}~\bibnamefont
  {Amet}}, \bibinfo {author} {\bibfnamefont {A.~J.}\ \bibnamefont {Bestwick}},
  \bibinfo {author} {\bibfnamefont {J.~R.}\ \bibnamefont {Williams}}, \bibinfo
  {author} {\bibfnamefont {L.}~\bibnamefont {Balicas}}, \bibinfo {author}
  {\bibfnamefont {K.}~\bibnamefont {Watanabe}}, \bibinfo {author}
  {\bibfnamefont {T.}~\bibnamefont {Taniguchi}}, \ and\ \bibinfo {author}
  {\bibfnamefont {D.}~\bibnamefont {Goldhaber-Gordon}},\ }\href {\doibase
  10.1038/ncomms6838} {\bibfield  {journal} {\bibinfo  {journal} {Nature
  Communications}\ }\textbf {\bibinfo {volume} {6}},\ \bibinfo {pages} {5838}
  (\bibinfo {year} {2015})}\BibitemShut {NoStop}%
\bibitem [{\citenamefont {Bestwick}\ \emph {et~al.}(2016)\citenamefont
  {Bestwick}, \citenamefont {Liang}, \citenamefont {Goldhaber-Gordon},
  \citenamefont {Amet}, \citenamefont {Diankov}, \citenamefont {Jaroszynski},
  \citenamefont {Watanabe}, \citenamefont {Tharratt}, \citenamefont {Lee},
  \citenamefont {Gallagher}, \citenamefont {Taniguchi},\ and\ \citenamefont
  {Coniglio}}]{bestwick_robust_2016}%
  \BibitemOpen
  \bibfield  {author} {\bibinfo {author} {\bibfnamefont {A.~J.}\ \bibnamefont
  {Bestwick}}, \bibinfo {author} {\bibfnamefont {C.-T.}\ \bibnamefont {Liang}},
  \bibinfo {author} {\bibfnamefont {D.}~\bibnamefont {Goldhaber-Gordon}},
  \bibinfo {author} {\bibfnamefont {F.}~\bibnamefont {Amet}}, \bibinfo {author}
  {\bibfnamefont {G.}~\bibnamefont {Diankov}}, \bibinfo {author} {\bibfnamefont
  {J.}~\bibnamefont {Jaroszynski}}, \bibinfo {author} {\bibfnamefont
  {K.}~\bibnamefont {Watanabe}}, \bibinfo {author} {\bibfnamefont
  {K.}~\bibnamefont {Tharratt}}, \bibinfo {author} {\bibfnamefont
  {M.}~\bibnamefont {Lee}}, \bibinfo {author} {\bibfnamefont {P.}~\bibnamefont
  {Gallagher}}, \bibinfo {author} {\bibfnamefont {T.}~\bibnamefont
  {Taniguchi}}, \ and\ \bibinfo {author} {\bibfnamefont {W.}~\bibnamefont
  {Coniglio}},\ }\href {\doibase 10.1038/ncomms13908} {\bibfield  {journal}
  {\bibinfo  {journal} {Nature Communications}\ }\textbf {\bibinfo {volume}
  {7}},\ \bibinfo {pages} {13908} (\bibinfo {year} {2016})}\BibitemShut
  {NoStop}%
\bibitem [{\citenamefont {Li}\ \emph {et~al.}(2017)\citenamefont {Li},
  \citenamefont {Tan}, \citenamefont {Chen}, \citenamefont {Zeng},
  \citenamefont {Taniguchi}, \citenamefont {Watanabe}, \citenamefont {Hone},\
  and\ \citenamefont {Dean}}]{li_even_2017}%
  \BibitemOpen
  \bibfield  {author} {\bibinfo {author} {\bibfnamefont {J.~I.~A.}\
  \bibnamefont {Li}}, \bibinfo {author} {\bibfnamefont {C.}~\bibnamefont
  {Tan}}, \bibinfo {author} {\bibfnamefont {S.}~\bibnamefont {Chen}}, \bibinfo
  {author} {\bibfnamefont {Y.}~\bibnamefont {Zeng}}, \bibinfo {author}
  {\bibfnamefont {T.}~\bibnamefont {Taniguchi}}, \bibinfo {author}
  {\bibfnamefont {K.}~\bibnamefont {Watanabe}}, \bibinfo {author}
  {\bibfnamefont {J.}~\bibnamefont {Hone}}, \ and\ \bibinfo {author}
  {\bibfnamefont {C.~R.}\ \bibnamefont {Dean}},\ }\href {\doibase
  10.1126/science.aao2521} {\bibfield  {journal} {\bibinfo  {journal}
  {Science}\ ,\ \bibinfo {pages} {eaao2521}} (\bibinfo {year}
  {2017})}\BibitemShut {NoStop}%
\bibitem [{\citenamefont {Zibrov}\ \emph
  {et~al.}(2017{\natexlab{b}})\citenamefont {Zibrov}, \citenamefont {Spanton},
  \citenamefont {Zhou}, \citenamefont {Kometter}, \citenamefont {Taniguchi},
  \citenamefont {Watanabe},\ and\ \citenamefont {Young}}]{zibrov_even_2017}%
  \BibitemOpen
  \bibfield  {author} {\bibinfo {author} {\bibfnamefont {A.~A.}\ \bibnamefont
  {Zibrov}}, \bibinfo {author} {\bibfnamefont {E.~M.}\ \bibnamefont {Spanton}},
  \bibinfo {author} {\bibfnamefont {H.}~\bibnamefont {Zhou}}, \bibinfo {author}
  {\bibfnamefont {C.}~\bibnamefont {Kometter}}, \bibinfo {author}
  {\bibfnamefont {T.}~\bibnamefont {Taniguchi}}, \bibinfo {author}
  {\bibfnamefont {K.}~\bibnamefont {Watanabe}}, \ and\ \bibinfo {author}
  {\bibfnamefont {A.~F.}\ \bibnamefont {Young}},\ }\href
  {http://arxiv.org/abs/1712.01968} {\bibfield  {journal} {\bibinfo  {journal}
  {arXiv:1712.01968 [cond-mat]}\ } (\bibinfo {year} {2017}{\natexlab{b}})},\
  \bibinfo {note} {arXiv: 1712.01968}\BibitemShut {NoStop}%
\bibitem [{\citenamefont {Yan}\ and\ \citenamefont {Fuhrer}(2010)}]{Yan_2010}%
  \BibitemOpen
  \bibfield  {author} {\bibinfo {author} {\bibfnamefont {J.}~\bibnamefont
  {Yan}}\ and\ \bibinfo {author} {\bibfnamefont {M.~S.}\ \bibnamefont
  {Fuhrer}},\ }\href {\doibase 10.1021/nl102459t} {\bibfield  {journal}
  {\bibinfo  {journal} {Nano Letters}\ }\textbf {\bibinfo {volume} {10}},\
  \bibinfo {pages} {4521} (\bibinfo {year} {2010})}\BibitemShut {NoStop}%
\bibitem [{\citenamefont {Zhu}\ \emph {et~al.}(2018)\citenamefont {Zhu},
  \citenamefont {Wang}, \citenamefont {Fu}, \citenamefont {Pfeiffer},
  \citenamefont {West}, \citenamefont {Du},\ and\ \citenamefont
  {Lin}}]{Zhu_2018}%
  \BibitemOpen
  \bibfield  {author} {\bibinfo {author} {\bibfnamefont {Z.}~\bibnamefont
  {Zhu}}, \bibinfo {author} {\bibfnamefont {P.}~\bibnamefont {Wang}}, \bibinfo
  {author} {\bibfnamefont {H.}~\bibnamefont {Fu}}, \bibinfo {author}
  {\bibfnamefont {L.}~\bibnamefont {Pfeiffer}}, \bibinfo {author}
  {\bibfnamefont {K.}~\bibnamefont {West}}, \bibinfo {author} {\bibfnamefont
  {R.-R.}\ \bibnamefont {Du}}, \ and\ \bibinfo {author} {\bibfnamefont
  {X.}~\bibnamefont {Lin}},\ }\href {\doibase 10.1016/j.physe.2017.09.007}
  {\bibfield  {journal} {\bibinfo  {journal} {Physica E: Low-dimensional
  Systems and Nanostructures}\ }\textbf {\bibinfo {volume} {95}},\ \bibinfo
  {pages} {1} (\bibinfo {year} {2018})}\BibitemShut {NoStop}%
\bibitem [{\citenamefont {Peters}\ \emph {et~al.}(2014)\citenamefont {Peters},
  \citenamefont {Giesbers}, \citenamefont {Burghard},\ and\ \citenamefont
  {Kern}}]{Peters_2014}%
  \BibitemOpen
  \bibfield  {author} {\bibinfo {author} {\bibfnamefont {E.~C.}\ \bibnamefont
  {Peters}}, \bibinfo {author} {\bibfnamefont {A.~J.~M.}\ \bibnamefont
  {Giesbers}}, \bibinfo {author} {\bibfnamefont {M.}~\bibnamefont {Burghard}},
  \ and\ \bibinfo {author} {\bibfnamefont {K.}~\bibnamefont {Kern}},\ }\href
  {\doibase 10.1063/1.4878396} {\bibfield  {journal} {\bibinfo  {journal}
  {Applied Physics Letters}\ }\textbf {\bibinfo {volume} {104}},\ \bibinfo
  {pages} {203109} (\bibinfo {year} {2014})}\BibitemShut {NoStop}%
\bibitem [{\citenamefont {Zhao}\ \emph {et~al.}(2012)\citenamefont {Zhao},
  \citenamefont {Cadden-Zimansky}, \citenamefont {Ghahari},\ and\ \citenamefont
  {Kim}}]{Zhao_2012}%
  \BibitemOpen
  \bibfield  {author} {\bibinfo {author} {\bibfnamefont {Y.}~\bibnamefont
  {Zhao}}, \bibinfo {author} {\bibfnamefont {P.}~\bibnamefont
  {Cadden-Zimansky}}, \bibinfo {author} {\bibfnamefont {F.}~\bibnamefont
  {Ghahari}}, \ and\ \bibinfo {author} {\bibfnamefont {P.}~\bibnamefont
  {Kim}},\ }\href {\doibase 10.1103/PhysRevLett.108.106804} {\bibfield
  {journal} {\bibinfo  {journal} {Phys. Rev. Lett.}\ }\textbf {\bibinfo
  {volume} {108}},\ \bibinfo {pages} {106804} (\bibinfo {year}
  {2012})}\BibitemShut {NoStop}%
\bibitem [{\citenamefont {Hunt}\ \emph {et~al.}(2013)\citenamefont {Hunt},
  \citenamefont {Sanchez-Yamagishi}, \citenamefont {Young}, \citenamefont
  {Yankowitz}, \citenamefont {LeRoy}, \citenamefont {Watanabe}, \citenamefont
  {Taniguchi}, \citenamefont {Moon}, \citenamefont {Koshino}, \citenamefont
  {Jarillo-Herrero},\ and\ \citenamefont {Ashoori}}]{hunt_massive_2013}%
  \BibitemOpen
  \bibfield  {author} {\bibinfo {author} {\bibfnamefont {B.}~\bibnamefont
  {Hunt}}, \bibinfo {author} {\bibfnamefont {J.~D.}\ \bibnamefont
  {Sanchez-Yamagishi}}, \bibinfo {author} {\bibfnamefont {A.~F.}\ \bibnamefont
  {Young}}, \bibinfo {author} {\bibfnamefont {M.}~\bibnamefont {Yankowitz}},
  \bibinfo {author} {\bibfnamefont {B.~J.}\ \bibnamefont {LeRoy}}, \bibinfo
  {author} {\bibfnamefont {K.}~\bibnamefont {Watanabe}}, \bibinfo {author}
  {\bibfnamefont {T.}~\bibnamefont {Taniguchi}}, \bibinfo {author}
  {\bibfnamefont {P.}~\bibnamefont {Moon}}, \bibinfo {author} {\bibfnamefont
  {M.}~\bibnamefont {Koshino}}, \bibinfo {author} {\bibfnamefont
  {P.}~\bibnamefont {Jarillo-Herrero}}, \ and\ \bibinfo {author} {\bibfnamefont
  {R.~C.}\ \bibnamefont {Ashoori}},\ }\href {\doibase 10.1126/science.1237240}
  {\bibfield  {journal} {\bibinfo  {journal} {Science}\ }\textbf {\bibinfo
  {volume} {340}},\ \bibinfo {pages} {1427} (\bibinfo {year}
  {2013})}\BibitemShut {NoStop}%
\bibitem [{\citenamefont {Amet}\ \emph {et~al.}(2013)\citenamefont {Amet},
  \citenamefont {Williams}, \citenamefont {Watanabe}, \citenamefont
  {Taniguchi},\ and\ \citenamefont {Goldhaber-Gordon}}]{amet_insulating_2013}%
  \BibitemOpen
  \bibfield  {author} {\bibinfo {author} {\bibfnamefont {F.}~\bibnamefont
  {Amet}}, \bibinfo {author} {\bibfnamefont {J.~R.}\ \bibnamefont {Williams}},
  \bibinfo {author} {\bibfnamefont {K.}~\bibnamefont {Watanabe}}, \bibinfo
  {author} {\bibfnamefont {T.}~\bibnamefont {Taniguchi}}, \ and\ \bibinfo
  {author} {\bibfnamefont {D.}~\bibnamefont {Goldhaber-Gordon}},\ }\href
  {\doibase 10.1103/PhysRevLett.110.216601} {\bibfield  {journal} {\bibinfo
  {journal} {Physical Review Letters}\ }\textbf {\bibinfo {volume} {110}},\
  \bibinfo {pages} {216601} (\bibinfo {year} {2013})}\BibitemShut {NoStop}%
\bibitem [{\citenamefont {Morf}\ \emph {et~al.}(2002)\citenamefont {Morf},
  \citenamefont {d{\textquoteright}Ambrumenil},\ and\ \citenamefont
  {Das~Sarma}}]{morf_excitation_2002}%
  \BibitemOpen
  \bibfield  {author} {\bibinfo {author} {\bibfnamefont {R.~H.}\ \bibnamefont
  {Morf}}, \bibinfo {author} {\bibfnamefont {N.}~\bibnamefont
  {d{\textquoteright}Ambrumenil}}, \ and\ \bibinfo {author} {\bibfnamefont
  {S.}~\bibnamefont {Das~Sarma}},\ }\href {\doibase 10.1103/PhysRevB.66.075408}
  {\bibfield  {journal} {\bibinfo  {journal} {Physical Review B}\ }\textbf
  {\bibinfo {volume} {66}},\ \bibinfo {pages} {075408} (\bibinfo {year}
  {2002})}\BibitemShut {NoStop}%
\bibitem [{\citenamefont {Abanin}\ \emph {et~al.}(2013)\citenamefont {Abanin},
  \citenamefont {Feldman}, \citenamefont {Yacoby},\ and\ \citenamefont
  {Halperin}}]{abanin_fractional_2013}%
  \BibitemOpen
  \bibfield  {author} {\bibinfo {author} {\bibfnamefont {D.~A.}\ \bibnamefont
  {Abanin}}, \bibinfo {author} {\bibfnamefont {B.~E.}\ \bibnamefont {Feldman}},
  \bibinfo {author} {\bibfnamefont {A.}~\bibnamefont {Yacoby}}, \ and\ \bibinfo
  {author} {\bibfnamefont {B.~I.}\ \bibnamefont {Halperin}},\ }\href {\doibase
  10.1103/PhysRevB.88.115407} {\bibfield  {journal} {\bibinfo  {journal}
  {Physical Review B}\ }\textbf {\bibinfo {volume} {88}},\ \bibinfo {pages}
  {115407} (\bibinfo {year} {2013})}\BibitemShut {NoStop}%
\bibitem [{\citenamefont {Dethlefsen}\ \emph {et~al.}(2006)\citenamefont
  {Dethlefsen}, \citenamefont {Mariani}, \citenamefont {Tranitz}, \citenamefont
  {Wegscheider},\ and\ \citenamefont {Haug}}]{dethlefsen_signatures_2006_b}%
  \BibitemOpen
  \bibfield  {author} {\bibinfo {author} {\bibfnamefont {A.~F.}\ \bibnamefont
  {Dethlefsen}}, \bibinfo {author} {\bibfnamefont {E.}~\bibnamefont {Mariani}},
  \bibinfo {author} {\bibfnamefont {H.-P.}\ \bibnamefont {Tranitz}}, \bibinfo
  {author} {\bibfnamefont {W.}~\bibnamefont {Wegscheider}}, \ and\ \bibinfo
  {author} {\bibfnamefont {R.~J.}\ \bibnamefont {Haug}},\ }\href {\doibase
  10.1103/PhysRevB.74.165325} {\bibfield  {journal} {\bibinfo  {journal} {Phys.
  Rev. B}\ }\textbf {\bibinfo {volume} {74}},\ \bibinfo {pages} {165325}
  (\bibinfo {year} {2006})}\BibitemShut {NoStop}%
\bibitem [{\citenamefont {Geick}\ \emph {et~al.}(1966)\citenamefont {Geick},
  \citenamefont {Perry},\ and\ \citenamefont {Rupprecht}}]{geick_normal_1966}%
  \BibitemOpen
  \bibfield  {author} {\bibinfo {author} {\bibfnamefont {R.}~\bibnamefont
  {Geick}}, \bibinfo {author} {\bibfnamefont {C.~H.}\ \bibnamefont {Perry}}, \
  and\ \bibinfo {author} {\bibfnamefont {G.}~\bibnamefont {Rupprecht}},\ }\href
  {\doibase 10.1103/PhysRev.146.543} {\bibfield  {journal} {\bibinfo  {journal}
  {Physical Review}\ }\textbf {\bibinfo {volume} {146}},\ \bibinfo {pages}
  {543} (\bibinfo {year} {1966})}\BibitemShut {NoStop}%
\bibitem [{\citenamefont {Balram}\ \emph
  {et~al.}(2015{\natexlab{a}})\citenamefont {Balram}, \citenamefont {Toke},
  \citenamefont {Wojs},\ and\ \citenamefont {Jain}}]{balram_phase_2015}%
  \BibitemOpen
  \bibfield  {author} {\bibinfo {author} {\bibfnamefont {A.~C.}\ \bibnamefont
  {Balram}}, \bibinfo {author} {\bibfnamefont {C.}~\bibnamefont {Toke}},
  \bibinfo {author} {\bibfnamefont {A.}~\bibnamefont {Wojs}}, \ and\ \bibinfo
  {author} {\bibfnamefont {J.~K.}\ \bibnamefont {Jain}},\ }\href {\doibase
  10.1103/PhysRevB.91.045109} {\bibfield  {journal} {\bibinfo  {journal}
  {Physical Review B}\ }\textbf {\bibinfo {volume} {91}},\ \bibinfo {pages}
  {045109} (\bibinfo {year} {2015}{\natexlab{a}})}\BibitemShut {NoStop}%
\bibitem [{\citenamefont {Mandal}\ and\ \citenamefont
  {Jain}(2001)}]{mandal_low-energy_2001}%
  \BibitemOpen
  \bibfield  {author} {\bibinfo {author} {\bibfnamefont {S.~S.}\ \bibnamefont
  {Mandal}}\ and\ \bibinfo {author} {\bibfnamefont {J.~K.}\ \bibnamefont
  {Jain}},\ }\href {\doibase 10.1103/PhysRevB.63.201310} {\bibfield  {journal}
  {\bibinfo  {journal} {Physical Review B}\ }\textbf {\bibinfo {volume} {63}},\
  \bibinfo {pages} {201310} (\bibinfo {year} {2001})}\BibitemShut {NoStop}%
\bibitem [{\citenamefont {Wurstbauer}\ \emph {et~al.}(2011)\citenamefont
  {Wurstbauer}, \citenamefont {Majumder}, \citenamefont {Mandal}, \citenamefont
  {Dujovne}, \citenamefont {Rhone}, \citenamefont {Dennis}, \citenamefont
  {Rigosi}, \citenamefont {Jain}, \citenamefont {Pinczuk}, \citenamefont
  {West},\ and\ \citenamefont {Pfeiffer}}]{wurstbauer_observation_2011}%
  \BibitemOpen
  \bibfield  {author} {\bibinfo {author} {\bibfnamefont {U.}~\bibnamefont
  {Wurstbauer}}, \bibinfo {author} {\bibfnamefont {D.}~\bibnamefont
  {Majumder}}, \bibinfo {author} {\bibfnamefont {S.~S.}\ \bibnamefont
  {Mandal}}, \bibinfo {author} {\bibfnamefont {I.}~\bibnamefont {Dujovne}},
  \bibinfo {author} {\bibfnamefont {T.~D.}\ \bibnamefont {Rhone}}, \bibinfo
  {author} {\bibfnamefont {B.~S.}\ \bibnamefont {Dennis}}, \bibinfo {author}
  {\bibfnamefont {A.~F.}\ \bibnamefont {Rigosi}}, \bibinfo {author}
  {\bibfnamefont {J.~K.}\ \bibnamefont {Jain}}, \bibinfo {author}
  {\bibfnamefont {A.}~\bibnamefont {Pinczuk}}, \bibinfo {author} {\bibfnamefont
  {K.~W.}\ \bibnamefont {West}}, \ and\ \bibinfo {author} {\bibfnamefont
  {L.~N.}\ \bibnamefont {Pfeiffer}},\ }\href {\doibase
  10.1103/PhysRevLett.107.066804} {\bibfield  {journal} {\bibinfo  {journal}
  {Physical Review Letters}\ }\textbf {\bibinfo {volume} {107}},\ \bibinfo
  {pages} {066804} (\bibinfo {year} {2011})}\BibitemShut {NoStop}%
\bibitem [{\citenamefont {Balram}\ \emph
  {et~al.}(2015{\natexlab{b}})\citenamefont {Balram}, \citenamefont
  {Wurstbauer}, \citenamefont {W{\'o}js}, \citenamefont {Pinczuk},\ and\
  \citenamefont {Jain}}]{balram_fractionally_2015}%
  \BibitemOpen
  \bibfield  {author} {\bibinfo {author} {\bibfnamefont {A.~C.}\ \bibnamefont
  {Balram}}, \bibinfo {author} {\bibfnamefont {U.}~\bibnamefont {Wurstbauer}},
  \bibinfo {author} {\bibfnamefont {A.}~\bibnamefont {W{\'o}js}}, \bibinfo
  {author} {\bibfnamefont {A.}~\bibnamefont {Pinczuk}}, \ and\ \bibinfo
  {author} {\bibfnamefont {J.~K.}\ \bibnamefont {Jain}},\ }\href {\doibase
  10.1038/ncomms9981} {\bibfield  {journal} {\bibinfo  {journal} {Nature
  Communications}\ }\textbf {\bibinfo {volume} {6}},\ \bibinfo {pages} {8981}
  (\bibinfo {year} {2015}{\natexlab{b}})}\BibitemShut {NoStop}%
\bibitem [{\citenamefont {V{\'y}born{\'y}}\ \emph {et~al.}(2009)\citenamefont
  {V{\'y}born{\'y}}, \citenamefont {Dethlefsen}, \citenamefont {Haug},\ and\
  \citenamefont {W{\'o}js}}]{vyborny_charge-spin_2009}%
  \BibitemOpen
  \bibfield  {author} {\bibinfo {author} {\bibfnamefont {K.}~\bibnamefont
  {V{\'y}born{\'y}}}, \bibinfo {author} {\bibfnamefont {A.~F.}\ \bibnamefont
  {Dethlefsen}}, \bibinfo {author} {\bibfnamefont {R.~J.}\ \bibnamefont
  {Haug}}, \ and\ \bibinfo {author} {\bibfnamefont {A.}~\bibnamefont
  {W{\'o}js}},\ }\href {\doibase 10.1103/PhysRevB.80.045407} {\bibfield
  {journal} {\bibinfo  {journal} {Physical Review B}\ }\textbf {\bibinfo
  {volume} {80}},\ \bibinfo {pages} {045407} (\bibinfo {year}
  {2009})}\BibitemShut {NoStop}%
\bibitem [{\citenamefont {Geraedts}\ \emph {et~al.}(2015)\citenamefont
  {Geraedts}, \citenamefont {Zaletel}, \citenamefont {Papi{\'c}},\ and\
  \citenamefont {Mong}}]{geraedts_competing_2015}%
  \BibitemOpen
  \bibfield  {author} {\bibinfo {author} {\bibfnamefont {S.}~\bibnamefont
  {Geraedts}}, \bibinfo {author} {\bibfnamefont {M.~P.}\ \bibnamefont
  {Zaletel}}, \bibinfo {author} {\bibfnamefont {Z.}~\bibnamefont {Papi{\'c}}},
  \ and\ \bibinfo {author} {\bibfnamefont {R.~S.~K.}\ \bibnamefont {Mong}},\
  }\href {\doibase 10.1103/PhysRevB.91.205139} {\bibfield  {journal} {\bibinfo
  {journal} {Physical Review B}\ }\textbf {\bibinfo {volume} {91}},\ \bibinfo
  {pages} {205139} (\bibinfo {year} {2015})}\BibitemShut {NoStop}%
\bibitem [{SI()}]{SI}%
  \BibitemOpen
  \href@noop {} {\ }\bibinfo {note} {See the supplementary
  materials.}\BibitemShut {Stop}%
\bibitem [{\citenamefont {Alicea}\ and\ \citenamefont
  {Fisher}(2006)}]{alicea_graphene_2006_b}%
  \BibitemOpen
  \bibfield  {author} {\bibinfo {author} {\bibfnamefont {J.}~\bibnamefont
  {Alicea}}\ and\ \bibinfo {author} {\bibfnamefont {M.~P.~A.}\ \bibnamefont
  {Fisher}},\ }\href {\doibase 10.1103/PhysRevB.74.075422} {\bibfield
  {journal} {\bibinfo  {journal} {Phys. Rev. B}\ }\textbf {\bibinfo {volume}
  {74}},\ \bibinfo {pages} {075422} (\bibinfo {year} {2006})}\BibitemShut
  {NoStop}%
\bibitem [{\citenamefont {Kharitonov}(2012)}]{kharitonov_phase_2012}%
  \BibitemOpen
  \bibfield  {author} {\bibinfo {author} {\bibfnamefont {M.}~\bibnamefont
  {Kharitonov}},\ }\href {http://link.aps.org/doi/10.1103/PhysRevB.85.155439}
  {\bibfield  {journal} {\bibinfo  {journal} {Phys. Rev. B}\ }\textbf {\bibinfo
  {volume} {85}},\ \bibinfo {pages} {155439} (\bibinfo {year}
  {2012})}\BibitemShut {NoStop}%
\bibitem [{\citenamefont {Alicea}\ and\ \citenamefont
  {Fendley}(2016)}]{alicea_topological_2016}%
  \BibitemOpen
  \bibfield  {author} {\bibinfo {author} {\bibfnamefont {J.}~\bibnamefont
  {Alicea}}\ and\ \bibinfo {author} {\bibfnamefont {P.}~\bibnamefont
  {Fendley}},\ }\href {\doibase 10.1146/annurev-conmatphys-031115-011336}
  {\bibfield  {journal} {\bibinfo  {journal} {Annual Review of Condensed Matter
  Physics}\ }\textbf {\bibinfo {volume} {7}},\ \bibinfo {pages} {119} (\bibinfo
  {year} {2016})}\BibitemShut {NoStop}%
\end{thebibliography}
%

\pagebreak
\widetext
\begin{center}

\textbf{Supplementary information for: Quantitative transport measurements of fractional quantum Hall energy gaps in edgeless graphene devices}
\end{center}

\setcounter{equation}{0}
\setcounter{figure}{0}
\setcounter{table}{0}
\setcounter{page}{1}
\makeatletter
\renewcommand{\theequation}{S\arabic{equation}}
\renewcommand{\thefigure}{S\arabic{figure}}
\renewcommand{\bibnumfmt}[1]{[S#1]}
\renewcommand{\citenumfont}[1]{S#1}

\section{Fabrication and measurement details}

\begin{figure*}[b]
\includegraphics[]{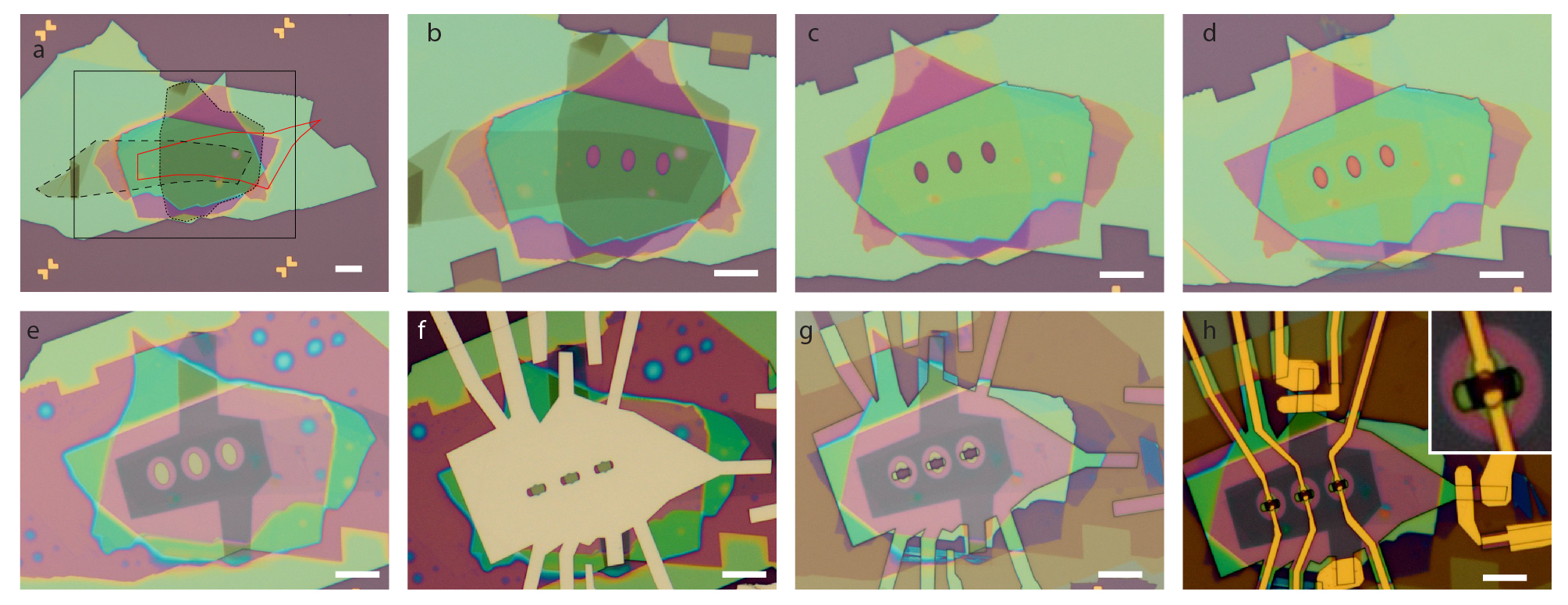}
\caption{Fabrication steps for Device A.
(a) Unprocessed stack. Dashed, dotted  and solid lines mark the outlines of  the bottom gate flake, top gate flake and graphene flakes in the stack.  Rectangle marks the size of the region shown in (b)-(h).
(b) The stack after partial etch which shaped the top graphite flake (bottom gate of the completed device).
(c) The stack after flipping.
(d) After shaping the top gate with another partial etch.
(e) Stack is covered with hBN flake.
(f) Lift-off is used to make an Al mask for the final etch which   shapes the device and opens apertures for edge contacts to graphene layer.
(g) After the final etch.
(h) Completed device after the deposition of Au/Pd/Cr contacts. Inset shows the structure of a contact region in Device A. Two green regions adjacent to edge contacts are substrate gated regions of MLG. Purple region is gated by bottom gate only. Dark region is gated by both top and bottom gates of the device.
 All scalebars in (a)-(h) correspond to 10~$\mu \text{m} $.}
\label{FabricationSteps}
\end{figure*}

 Optical images of the different fabrication steps of Device A are shown in Fig.~\ref{FabricationSteps}. The initial stack had the following structure: hBN/graphite/hBN/MLG/hBN/graphite (top-to-bottom).  hBN flakes used in devices A and B have thickness of 50-60~nm. The openings for contact regions were etched in the top hBN/graphite layers by a partial  $\text{CHF}_3$/$\text{O}_2$  etch with PMMA mask. Then the stack was picked up from substrate onto PPC/PDMS structure analogous to those used during the stack assembly. PPC film was separated from the PDMS substrate and applied to a new substrate so that the stack is on top of the film. Then the PPC between the stack and the substrate was removed by annealing at 375$^\circ$~C. This accomplishes flipping the stack (Fig.~~\ref{FabricationSteps}(c)).  $\text{O}_2$ etch is used to make openings in the top graphite layer, which are aligned with the openings in the bottom gate but have a larger size. The partial stack is them covered with another hBN flake to isolate the top gate from the contact leads which run on the top surface of the device. A 40~nm thick Al mask made by lift-off was used for the final etch  to define the shape of the device and to etch tranches in the contact regions where the MLG layer is contacted. The contacts were made by deposition of Cr/Pd/Au contacts 3/15/150~nm  and subsequent lift off.

Device A has three islands where the contact to the dual-graphite-gated MLG is made. One of them is shown in the inset of Fig.~\ref{FabricationSteps}(h). The center of each island has a trench which is etched through the entire stack. Adjacent to each side of the trench are two small contact regions, in which graphene is gated only by the substrate. Voltage of -20 to -70~V is applied  to the substrate  during the measurement to dope the contact regions and decrease the contact resistance. The tranche and two  contact regions are surrounded by region of the stack where graphene is gated by the bottom gate. During the measurements of conductance in Device A, presented in this paper, two contacts belonging to different islands were used as a source and a drain. Bottom gate was used to control the carrier density in MLG layer, while the top gate was grounded.
In Device B, the contact  regions have simpler structure (inset in Fig.~\ref{CorbinoDevice2}(c)): nearly coincident circular windows are etched in both gates across which a trench is etched. During  the measurements top gate was used to dope the MLG layer and the bottom gate was grounded.

Most of the measurements were done in a dilution refrigerator equipped with a rotator probe. The measurements at high fields (\textgreater~ 14~T) were conducted in $\text{He}_3$ refrigerator at the National High Magnetic Field Lab. The conductance of the devices was measured by applying  ac voltage excitation of 50-200~$\mu \text{V}$ to one of the contacts which acted as a source. One of the other contacts served as a drain while the rest of the  contacts other than source and drain were left disconnect. The current flowing into the drain was measured by using current preamplifier(Ithaco) and lock in amplifier (SRS830).

 \section{Comparison of devices A and B}
\begin{figure*}[htb!]
\includegraphics[width=0.9\textwidth]{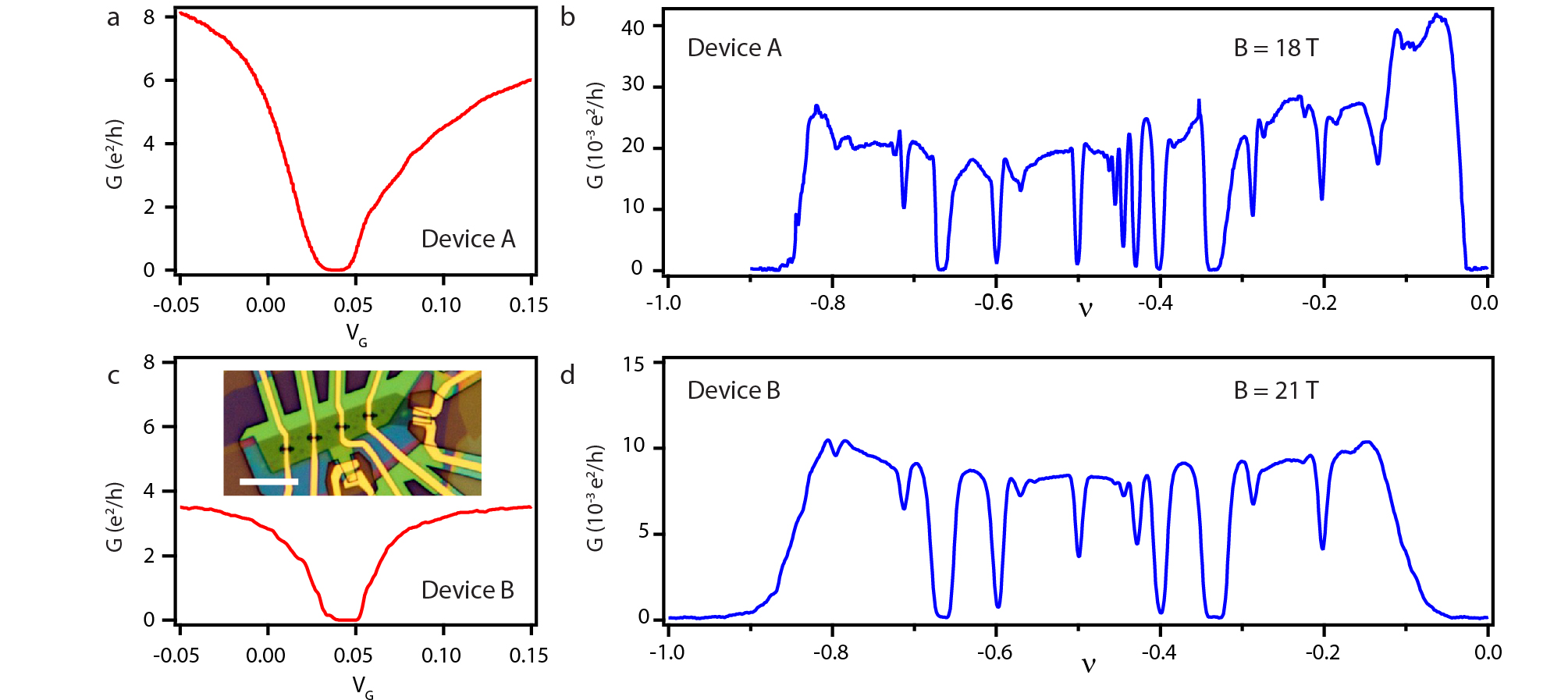}
\caption{Comparison of gaps in devices A and B.
(a),(c) Conductance  of device A and B accordingly, measured in zero magnetic field as a function of gate voltage. The measurement for Device A was done at 300~mK and for device B at 50~mK.
(b),(d) Half integer insulating states are observed for both devices A and B correspondingly at 18~T and 21~T.
The optical image of Device B is shown in the inset in (c). Scale bar is 10~$\mu \text{m}$}
\label{CorbinoDevice2}
\end{figure*}

 Both devices A and B have similar substrate induced sublattice splitting gap. It manifests itself as 20~mV insulating state measured at near the charge neutrality point (Fig.~\ref{CorbinoDevice2}(a), (c)). Further both devices show the insulating states at $\nu$=-1/2 which emerge at 18~T and 21~T accordingly in devices A and B (Fig.~\ref{CorbinoDevice2}(b),(d)).

 \section{FQH gaps and transitions}
\begin{figure*}[]
\includegraphics[width=0.9\textwidth]{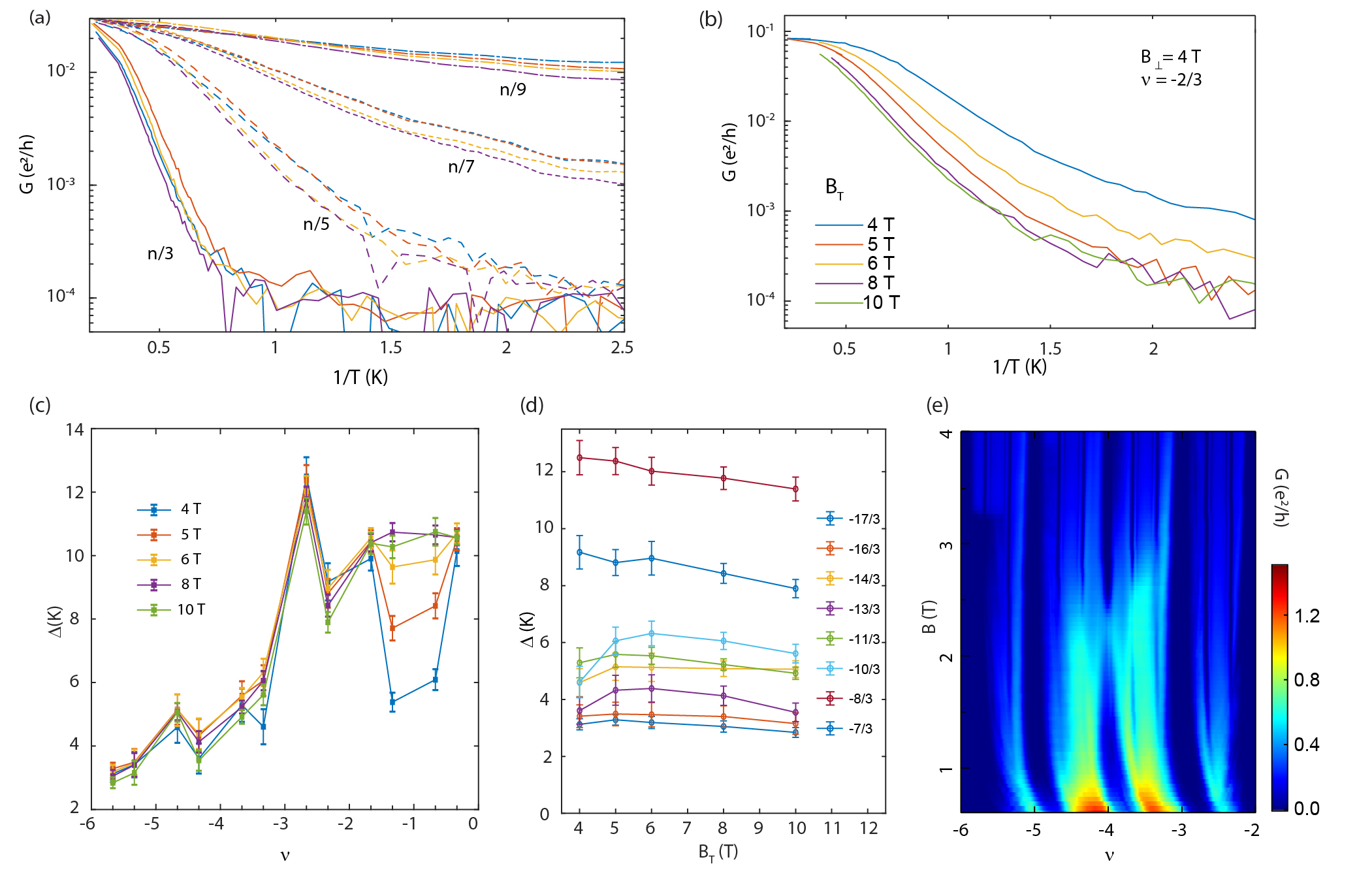}
\caption{FQH gaps and transitions in the first excited Landau level.
 (a) Conductance measurements showing thermally activated behavior.  The color coding of points corresponds to that of points shown in Fig.~2(c) in the main text.
 (b) Thermally activated behavior of conductance at $\nu=-2/3 $  at $B_{\perp}=4$~T and varying $B_{T}$. 
(c) n/3 gaps measured in device A at $B_{\perp}$=4~T as a function of $B_{T}$. While the gaps in $\nu$=-2/3, -4/3  FQH states show strong dependence on $B_{T}$, the gaps in rest of the $n/3$ FQH states remain nearly unchanged withing measurement error.
The dependence of $n/3$ in the first excited LL as a function of $B_{T}$  is shown separately in (d).
(e) Transition between insulating states at $\nu=-4$ in Device A is observed around 2.2~T.}
\label{FQHgaps}
\end{figure*}

 Figure~\ref{FQHgaps}(a)  shows thermally activated  behavior of conductance measured  for the $\nu=n/3$, $n/5$, $n/7$ and $n/9$ FQH states within the interval $\nu$=(-2,0)  at $B_\perp/B_T$=10T/14T  in Device~A. 
 Figure~\ref{FQHgaps}(b) shows thermally activated behavior of conductance at $\nu=-2/3 $  at $B_{\perp}=4$~T and varying $B_{T}$. 
 Activation  gaps measured in  $n/3$ FQH states for $\nu \in$~(-6, 0) at $B_{\perp}$=4~T and several values of $B_{T}$ are shown in Fig.~\ref{FQHgaps}(c). In contrast to  $\nu$=-2/3, -4/3 insulating states in which the gaps show strong dependence on $B_T$, the FQH states in the first excited Landau level remain nearly unchanged as $B_T$  increases (see also Fig.~\ref{FQHgaps}(d)). This suggest that the charge carrying excitations in $n/3$  gapped states for $\nu \in$~(-6,-4) states do not involve spin flip. Further, the relative magnitude of gaps measured at $B_{\perp}$=4~T is close to that measured at $B_{\perp}$=10~T. Similar to  Device B, the gap in the insulating state at $\nu$=-4  in Device A closes around 2.2~T.

\end{document}